\documentclass[a4paper,dvips 2t,superscriptaddress,showpacs]{revtex4}
\usepackage{epsfig,graphics,graphicx,amsmath,amssymb}

\usepackage{rotate}
\makeatletter
\def\myl{\mathopen\mybig}
\def\myr{\mathclose\mybig}
\def\mybigx#1{\dimen@#1\relax
        \mathchoice
        {\vbox to \dimen@{}}%
        {\vbox to \dimen@{}}%
        {\vbox to .7\dimen@{}}%
        {\vbox to .5\dimen@{}}}%

\def\mybig#1{{\hbox{$\left#1\mybigx{3.2em}\right.\n@space$}}}
\makeatother

\begin{document}
        
        \title{\bf\bf Coordinate space representation for  renormalization of quantum electrodynamics}
        \author{Amirhosein Mojavezi}\email{amojavezi98@gmail.com}
        \affiliation{Department of Physics, Ferdowsi University of Mashhad, 91775-1436 Mashhad, I.R. Iran}
        \author{Reza Moazzemi }\email{r.moazzemi@qom.ac.ir}\affiliation{Department of Physics, University of Qom, Ghadir Blvd., Qom 371614-6611, I.R. Iran} \author{Mohammad Ebrahim Zomorrodian}\affiliation{Department of Physics, Ferdowsi University of Mashhad, 91775-1436 Mashhad, I.R. Iran}

        \begin{abstract}
                In this paper we present a systematic treatment for fundamental renormalization of quantum electrodynamics in real space.  Although the standard renormalization is an old school problem in this case, it has not yet been completely done in position space. The most important difference with well-known differential renormalization is that we do the whole procedure in coordinate space without need to transformation to momentum space. Specially, we directly derive the conterterms in real space. This problem becomes important when the translational symmetry of the system breaks somehow explicitly (for example by nontrivial boundary condition (BC) on the fields). In this case, one is not able to move to momentum space by a simple Fourier  transformation. Therefore, in the context of renormalized perturbation theory, by imposing the renormalization conditions, counterterms in coordinate space will depend directly on the fields BCs (or background topology). Trivial BC or trivial background lead to the usual standard conterterms. If the counterterms modify then the quantum corrections of any physical quantity are different from those in free space where we have the translational invariance. We also show that, up to order $\alpha$, our counterterms are reduced to usual standard terms derived in free space.
        \end{abstract}
        
        \maketitle
        \section{Introduction}
        
        From its early stages, quantum field theory (QFT) was encountered some infinities leading to meaningless results and required to be eliminated. These ultraviolet (UV) infinities are related to the quantum corrections of some physical quantities, such as electron mass and charge \cite{A}. Very much attempts, starting with Kramers at 1940s \cite{B}, have been done  to control and remove these ultraviolet divergences (see for instance \cite{C}). In fact, to calculate a physical quantity (e.g. electron mass) in an interacting field theory, in addition to its `bare' value, we must take into account quantum corrections, $ \Delta m$:
        \begin{equation}
        {m_{\rm{physical}}} = {m_{\rm{bare}}} + \Delta m,
        \end{equation}
        where $ \Delta m$ is almost infinite  due to undetermined momenta in loop quantum corrections.
        
        Renormalization technique, is a recipe which consistently not only removes but also controls all infinities appeared in the theory  (for a qualitative review see for example\cite{D,E}). The importance of the
        renormalization procedure is not only to absorb divergences but also to complete the
        definition of the quantized field theory, i.e., the finite parts of the renormalization constants
        -fixed by the renormalization conditions- influence the results of the calculation of
        radiative corrections and therefore of physically observable quantities \cite{bohm}. In QFT, there are two completely equivalent methods for the systematic of renormalization; first, bare perturbation theory: working with the bare parameters and relate them to their physical values at the end of calculations. Roughly speaking,  the divergences are absorbed by \textit{redefinition} of unmeasurable bare quantities
        (see \cite{G,H,I}).  Second, renormalized perturbation theory: splitting the parameters appeared in the Lagrangian into two parts from beginning: physical part and \textit{counterterm} that absorbs unphysical part. In fact, the unobservable shifts between the bare and the physical parameters are absorbed by counterterms (see for instance \cite{J,K,L,M}).
        Both methods are required to give us precise definitions
        of the physical mass and coupling constants by applying renormalization conditions. The differences between two renormalization procedures are purely a matter of bookkeeping\cite{N,O,P}.
        
        There are many investigations related to renormalization programs concerned with quantum electrodynamics (QED)\cite{Q,R,S,T}, Quantum chromodynamics (QCD)\cite{U,V,W}, and scalar field with various self interactions (\cite{coleman,dashen,berzin1977,berzin1978}). All of these theories are renormalizable in four spacetime dimensions, since  their coupling  constants are dimensionless (Weinberg theorem)\cite{AA}. On the other hand, renormalization group (RG) methods have been vastly considered too(see for instance\cite{BB,CC,DD,EE}).

  We should do, in principle, the renormalization in position space. However, for the ease of calculation we do it in momentum space.  In fact,  there is a duality transformation from $p$- to $x$-space renormalization specially when we have  translational symmetry. One moves from position to momentum space by a simple Fourier transformation. This is easy to do if our wave functions are plain waves. But, if the translational symmetry breaks somehow explicitly, then the momentum is not a good quantum number and the wave functions are not plain waves, so that the transformation to momentum space is no longer so simple and trivial. In this case, field propagators will depend on nontrivial properties that break translational symmetry (e.g. nontrivial boundary conditions (BC) or nontrivial  background), therefore, all $n$-point functions and consequently all counterterms will depend on those nontrivial properties. (Please note that, it is not possible to remedy the renormalization in momentum space by any perturbation, since a nontrivial BC or a nonzero background is not a perturbative phenomenon \cite{PP}.) 
  
  We should here note that Differential Renormalization (DR) procedure \cite{freedman1,freedman2}, which has been vastly investigated in the literature, is done in coordinate space,  though the traditional method of renormalization in momentum space (for review see \cite{del,smirnov}). DR is equivalent to traditional renormalization \cite{pont,smirnov2,dunne}, and is based on the observation that the UV divergence reflects in the fact that the higher order amplitude cannot have a Fourier transform into momentum space due to the short-distance singularity. Thus one can, first, regulate such an amplitude by writing its singular parts as the derivatives of the normal functions, which have well defined Fourier transformation, and second, by performing the Fourier transformation in partial integration and discarding the surface term, directly get the renormalized result. In this procedure the surface terms which are dropped during the renormalization are just correspond to the counterterms. Therefore, to get the hidden counterterms we have to move to momentum space again. 

The derivation of standard counterterms from scattering amplitudes has been investigated from many years ago\cite{GG,HH,II,JJ}. In the context of DR there also exist some works in massive and massless QED \cite{haag1,haag2}. However, its program in position space has not yet been surveyed. 
In addition, the large order behavior of $\phi^4$ theory for nonzero background field is considered in \cite{parisi2017}. Also this theory in $1+1$ dimensions, renormalization in real space has been done in Ref. \cite{PP}.  Applications of the theory, where we have nontrivial BCs such as Dirichlet BC or nonzero background such az a kink have been used in Refs. \cite{LL} and \cite{OO}, respectively. In $3+1$ dimensions it has partially done in Ref. \cite{NN}.  In \cite{QQ} and \cite{RR} perturbative QFT in configuration space has been developed on curved space. Also, one can follow several recent works, for example, amplitudes in a massless QFT \cite{SS} and relativistic causality and position space renormalization \cite{TT}.
        
     In this paper, we shall systematically derive the countetrerms by imposing reasonable renormalization conditions in configuration space where there exist some nontrivial BC. As a matter of fact, the resultant counterterms should be equivalent to ones derived by standard renormalization in momentum space where we have a translational invariance. We will also present this in our paper.
        
        We have organized the paper as the following: We briefly review systematics of renormalization of QED theory in momentum space  in Sec. \ref{sec 2}.  Renormalized perturbation theory of QED as a systematic program in position space is considered in Sec. \ref{sec 3}. In Sec. \ref{sec 4} we compare our results with those in momentum space. Sec. \ref{sec 5}  summarizes our results and conclusions.
        
        \section{Renormalization of QED in momentum space: a brief review}
        \label{sec 2}
        In this section we  briefly  review systematics of renormalization for QED theory in momentum space (for complete details see \cite{GG}).  In general, any renormalizable QFT involves only a few superficially divergent amplitudes. In QED there are three amplitudes, involving four infinite constants; vertex correction \raisebox{-7mm}{\includegraphics[scale=.071]{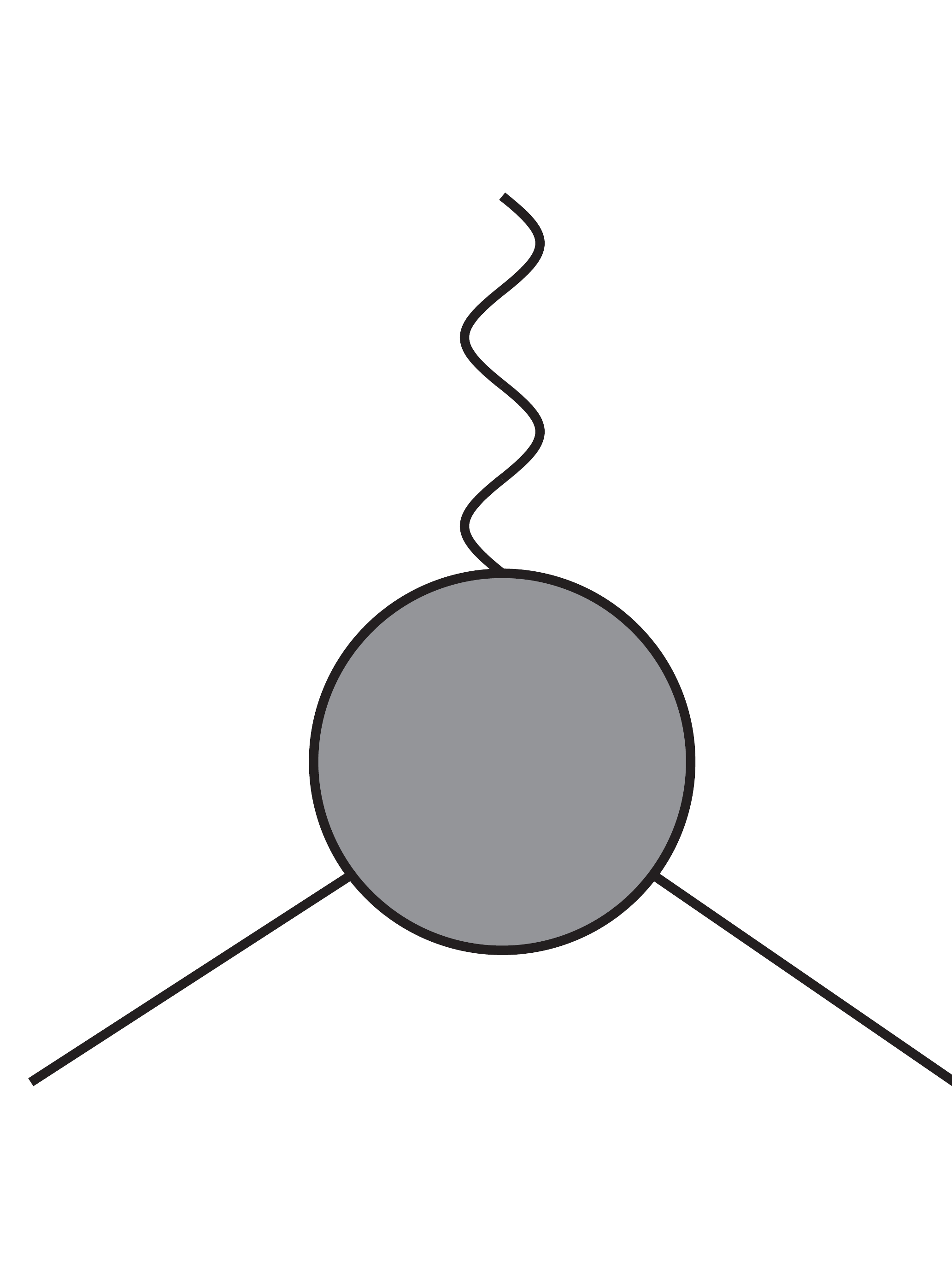}}, vacuum polarization  \raisebox{-9mm}{\includegraphics[scale=.071]{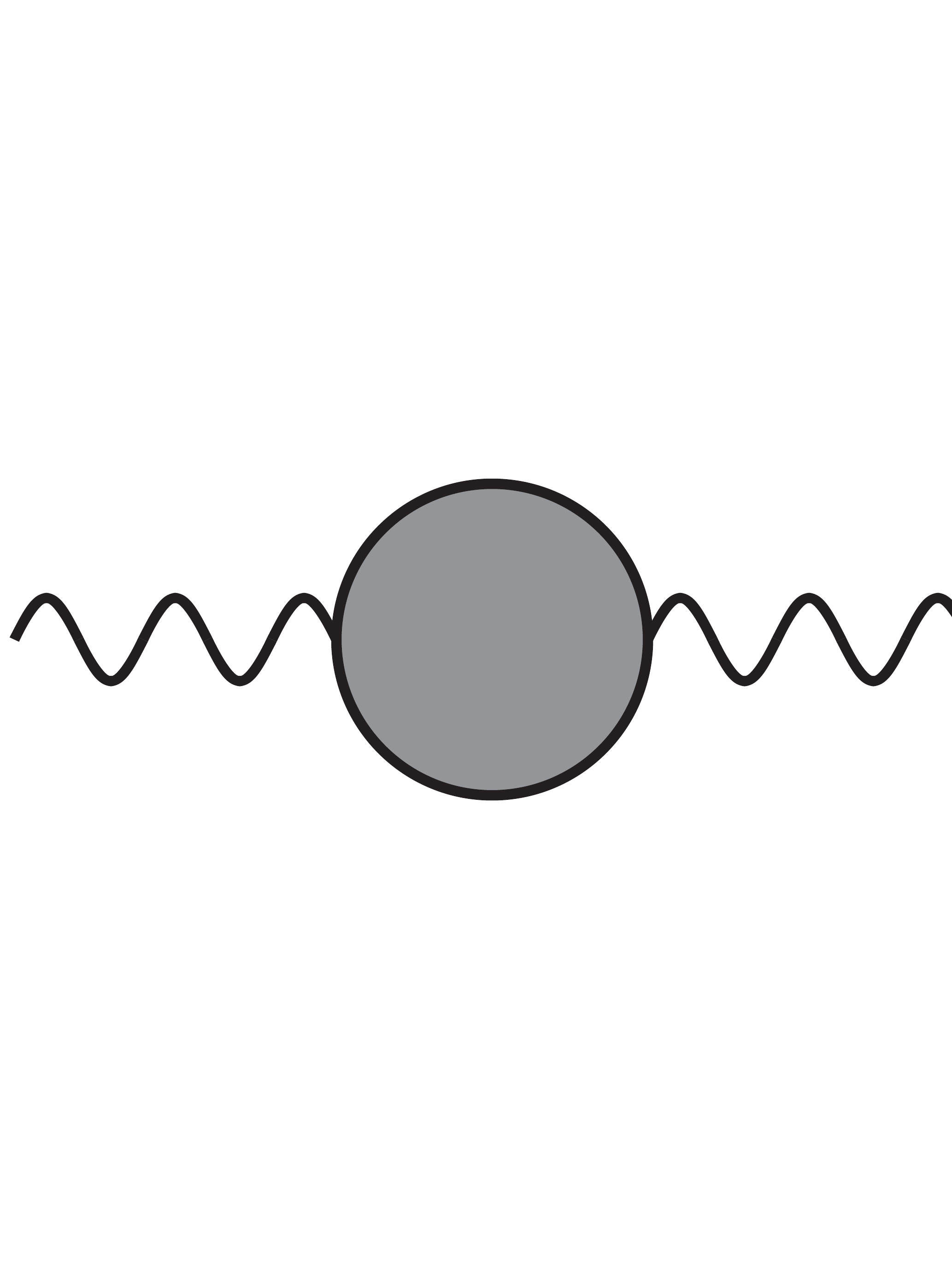}} and electron self energy  \raisebox{-9mm}{\includegraphics[scale=.07]{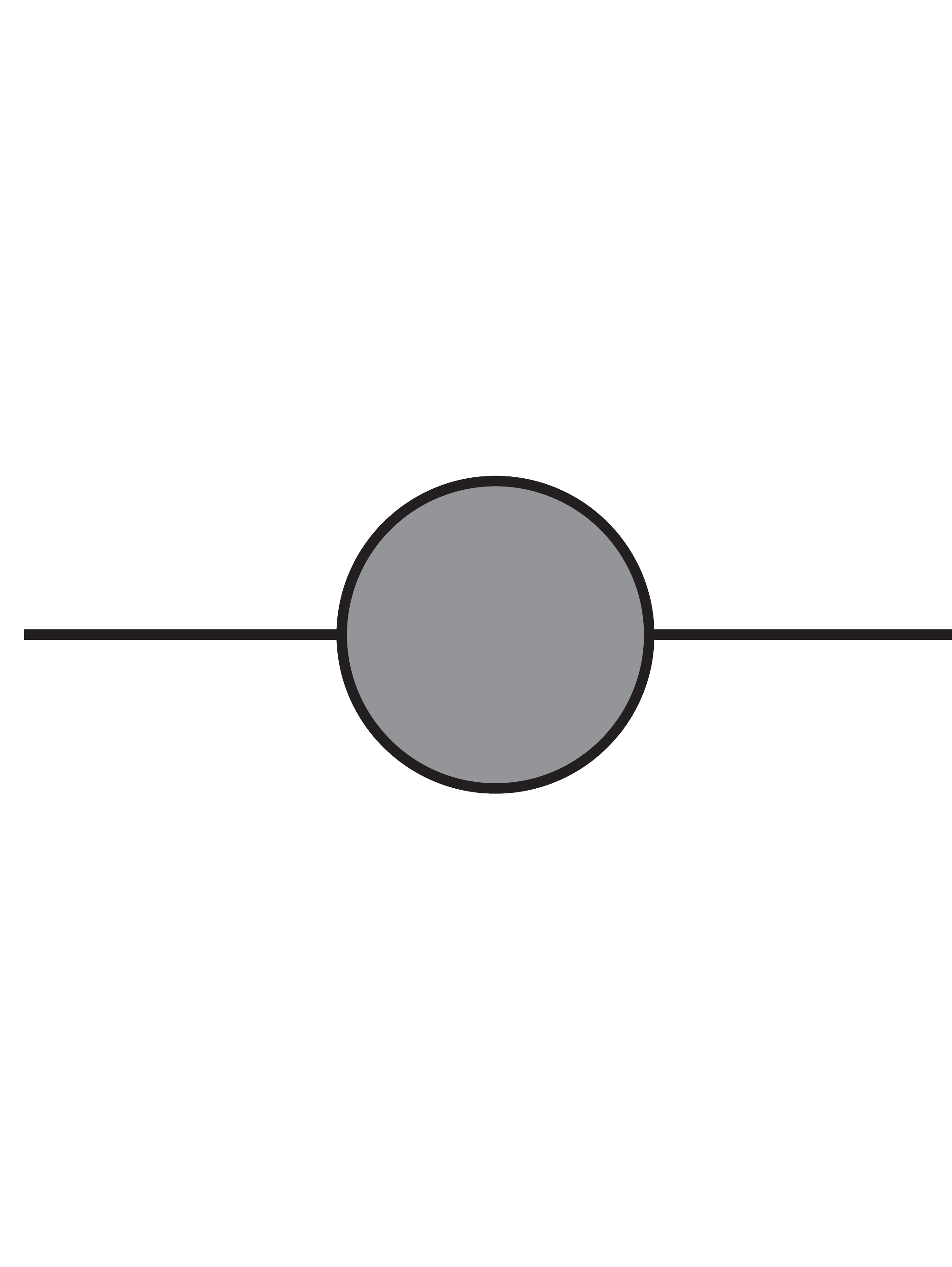}}.
        The aim of renormalized perturbation theory of QED is to absorb these constants into the four unobservable parameters of the theory: the bare mass, the bare coupling constant, the electron field strength and the photon field strength. The original QED Lagrangian is
        \begin{equation}\label{11}
        {\cal L_{\rm QED}}
        =-\frac{1}{4}(F^{\mu \nu})^2 +\bar{\Psi} (i\partial \hspace {-2mm}/-m_0)\Psi-e_0\bar{\Psi}\gamma_{\mu}\Psi A^\mu.
        \end{equation}
        where $m_0$ and $ e_0$ are the bare mass and the bare electric charge, respectively. The $\Psi(x)$ and $A^\mu(x)$ are fermion and photon fields, respectively, and can be written as 
        \begin{eqnarray}
        \Psi(x)&=&\int \frac{d^3\mathbf{p}}{(2\pi)^3}\sum\limits_{s=1,2}\frac{1}{\sqrt{2E_{\mathbf{p}}}}\left[c_{\mathbf{p}}^s\psi^s(x)+{d_{\mathbf{p}}^s}^\dag{\phi^s}(x)\right]
        \\A_\mu(x)&=&\int \frac{d^3\mathbf{p}}{(2\pi)^3}\sum\limits_{s=0}^{3}\frac{1}{\sqrt{2\omega_{\mathbf{p}}}}\left[a_{\mathbf{p}}^s\widetilde{A}^s_\mu(x)+{a_{\mathbf{p}}^s}^\dag{\widetilde{A}^{{s}^*}_\mu}(x)\right],
        \end{eqnarray}
        where, in the first line,  ${c_{\mathbf{p}}^s}^\dag$ ($c_{\mathbf{p}}^s$) and ${d_{\mathbf{p}}^s}^\dag$ ($d_{\mathbf{p}}^s$)  create (annihilate) a fermion and anti-fermion with momentum $\mathbf{p}$ and spin direction $s$, respectively. Here, $\psi^s(x)$ and ${\phi^s}(x)$ are the particle and anti-particle solutions of the Dirac equation, respectively. In the second line, ${a_{\mathbf{p}}^s}^\dag$ ($a_{\mathbf{p}}^s$) creates (annihilates) a photon with momentum $\mathbf{p}$ and polarization $\varepsilon_\mu^s(\mathbf{p})$, and $\widetilde{A}^r_\mu(x)$ are the momentum-space solution of the equation $\partial^\mu A_\mu=0$.
        
        By replacing $\Psi(x)= \sqrt{z_2}\Psi_r(x)$ and $A^{\mu}(x)=\sqrt{z_3}A^{\mu}_r(x) $, we have
        \begin{equation}\label{22}
        {\cal L_{\rm QED}}
        =-\frac{1}{4}z_3(F^{\mu \nu}_r)^2 +z_2 \bar{\Psi}_r (i\partial \hspace{-2mm}/ -m_0)\Psi_r-e_0z_2\sqrt{z_3}\bar{\Psi}_r\gamma_\mu\Psi_rA_{r}^\mu,
        \end{equation}
        where $z_2 $ and $ z_3 $ are the field-strength renormalizations for $ \Psi $ and $A^\mu $, respectively.
        We define a scaling factor $ z_1 $ as
        $
        ez_1= e_0z_2\sqrt{z_3}
        $
        and split each term of the Lagrangian into two pieces
        \begin{eqnarray}
        \nonumber{\cal L_{\rm QED}} =  &-& \frac{1}{4}{\left( {F_r^{\mu \nu }} \right)^2} + {\overline \Psi  _r}\left( {i\partial\hspace{-2mm}/ - m} \right){\Psi _r} - e{\overline \Psi  _r}{\gamma ^\mu }{\Psi _r}A_r^\mu \\ &-& \frac{1}{4}{\delta _3}{\left( {F_r^{\mu \nu }} \right)^2} + i{\delta _2}{\overline \Psi  _r}\partial\hspace{-2mm}/ {\Psi _r} - \left( {{\delta _m} + m{\delta _2}} \right){\overline \Psi  _r}{\Psi _r} - e{\delta _1}{\overline \Psi  _r}{\gamma _\mu }{\Psi _r}A_r^\mu ,\label{Lagrangian}
        \end{eqnarray}\\
        with $z_3=1+\delta_3$, $z_2=1+\delta_2$, $m_0=m+\delta_m$ and $ z_1=1+\delta_1$, where $\delta_1$, $\delta_2$, $\delta_3$ and $\delta_m$ are counterterms. Here, $m$ and $e$ are the physical mass and physical charge of the electron measured at large distances. Now, the Feynman rules for the above Lagrangian are:
        \begin{eqnarray}\label{44}
        \raisebox{-7mm}{\includegraphics[scale=.071]{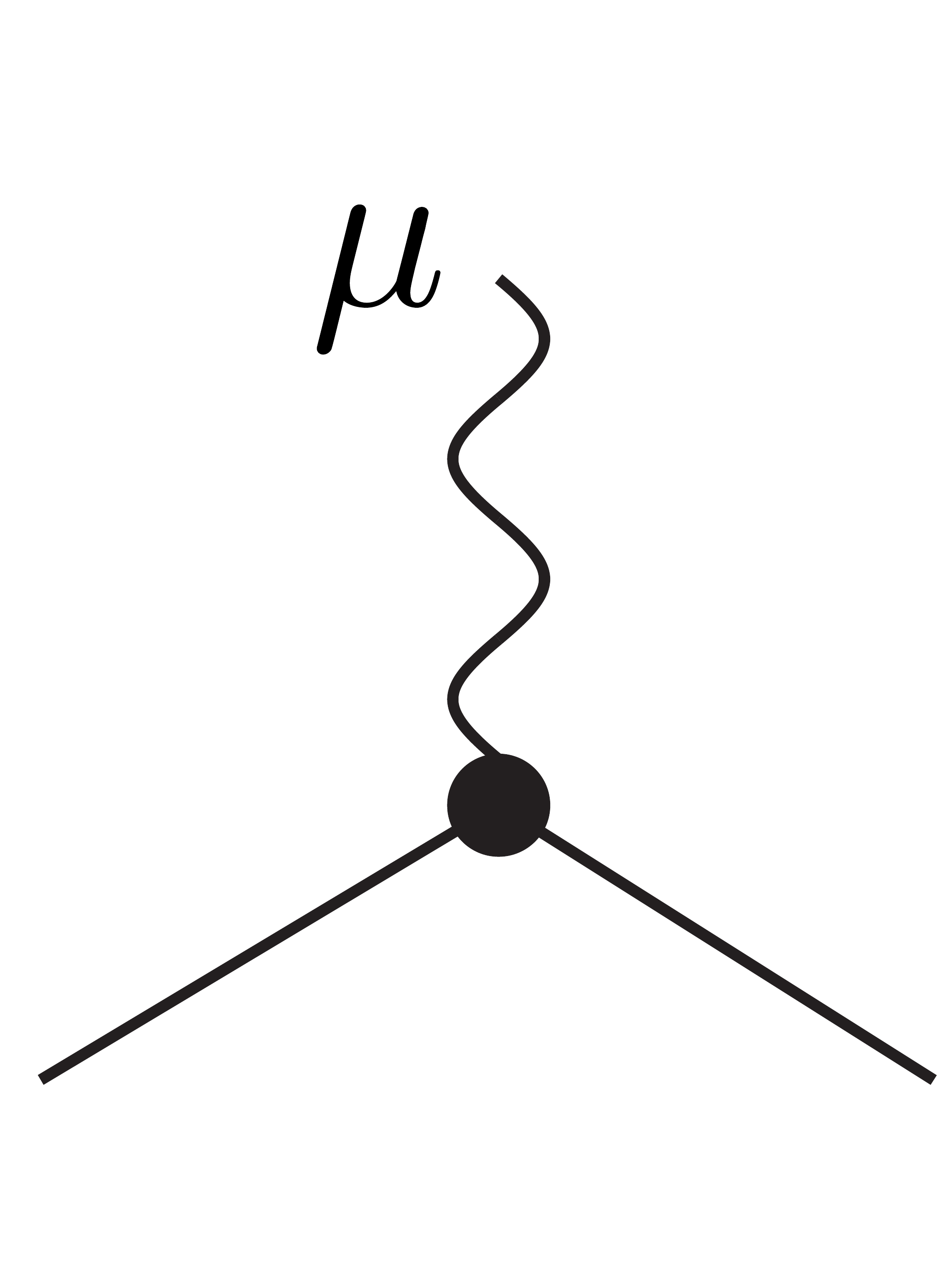}}&=&-ie\gamma ^{\mu}
        \\
        \raisebox{-6mm}{\includegraphics[scale=.071]{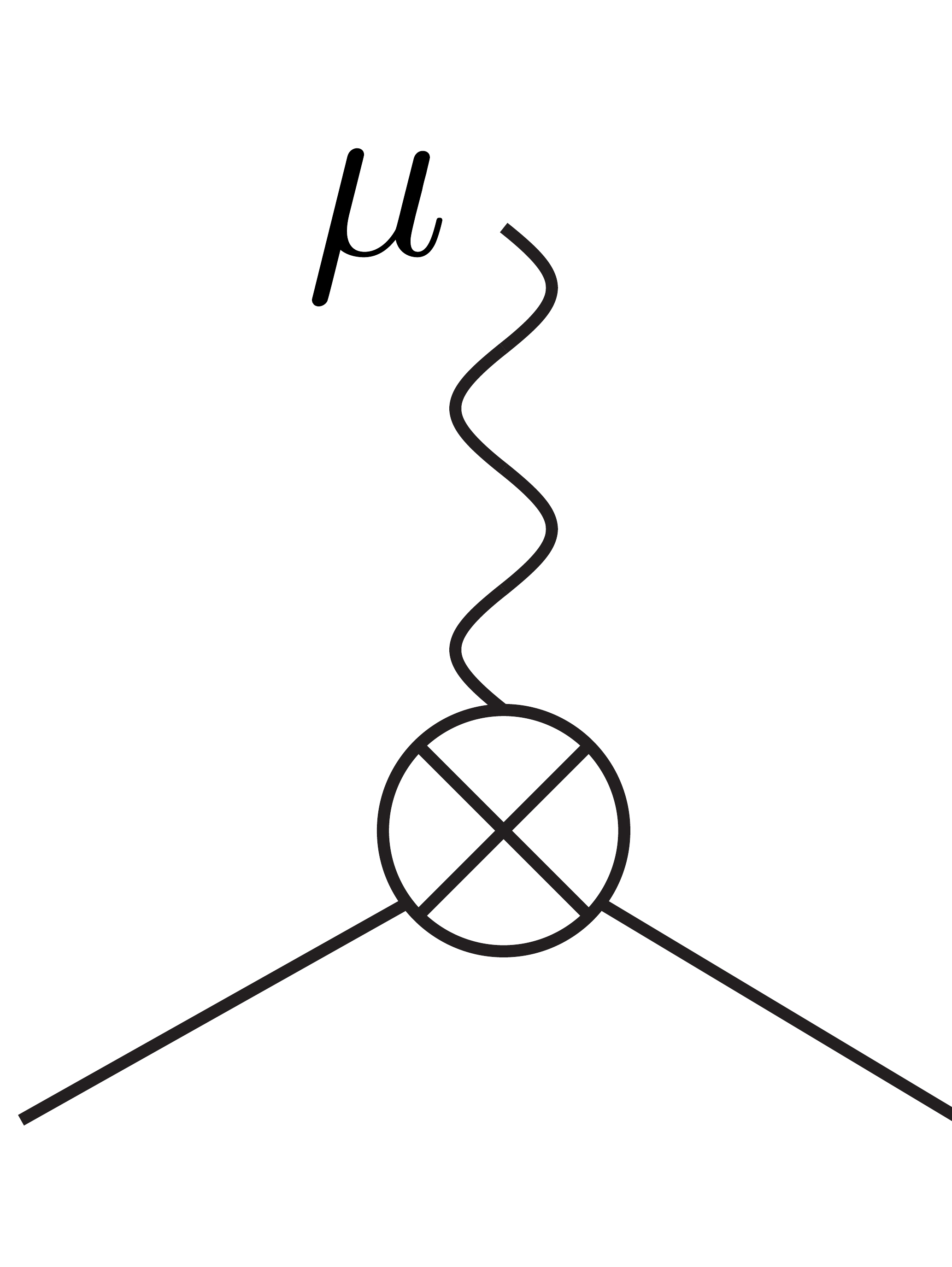}}&=& -ie\delta_1\gamma ^{\mu}
        \\\label{photon}
        \raisebox{-15mm}{\includegraphics[scale=.1]{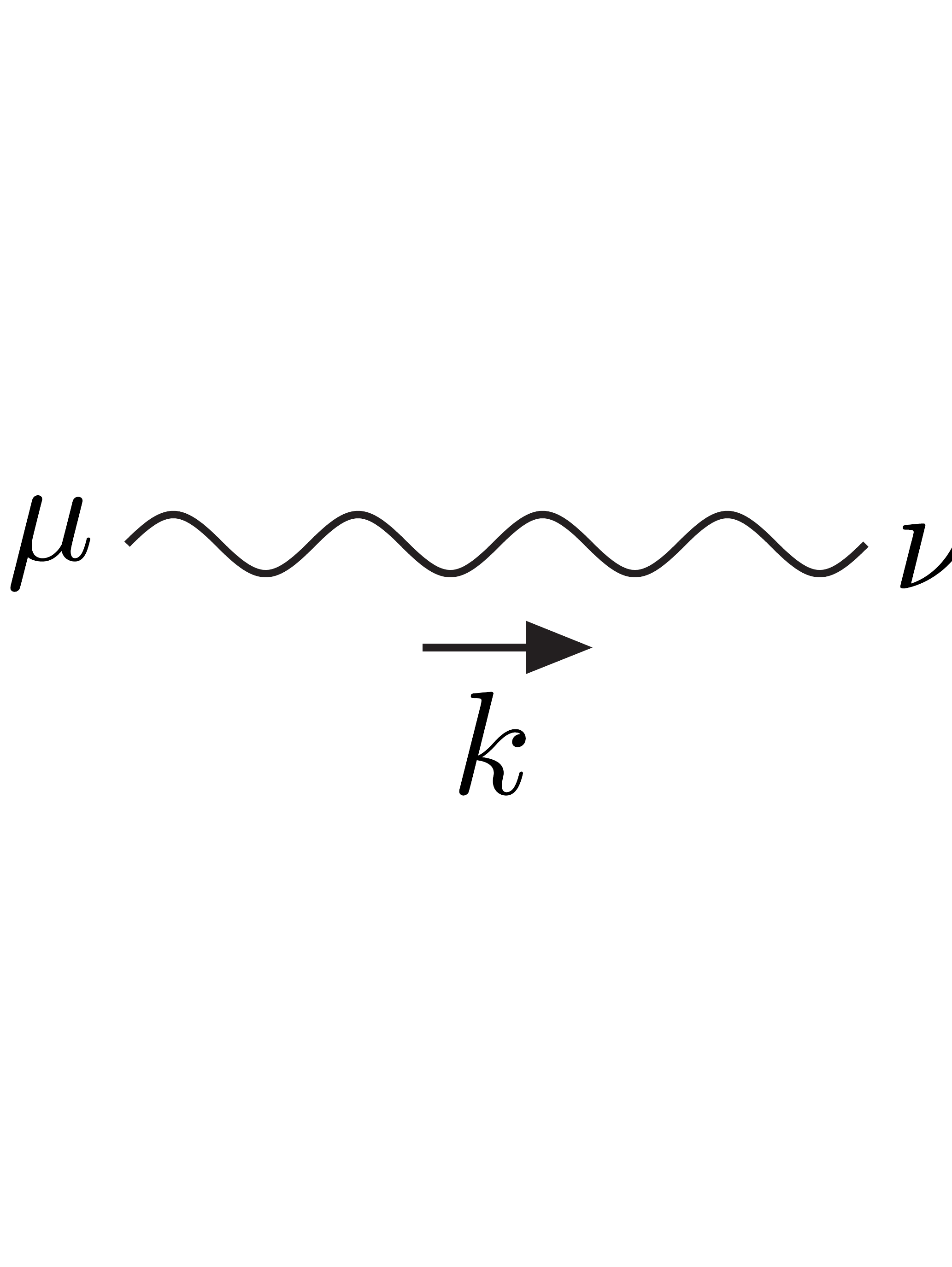}} &=& \frac{-ig^{\mu\nu}}{q^2+i\epsilon}   \text{ (Feynman gauge})
        \\\label{phver}
        \raisebox{-13mm}{\includegraphics[scale=.1]{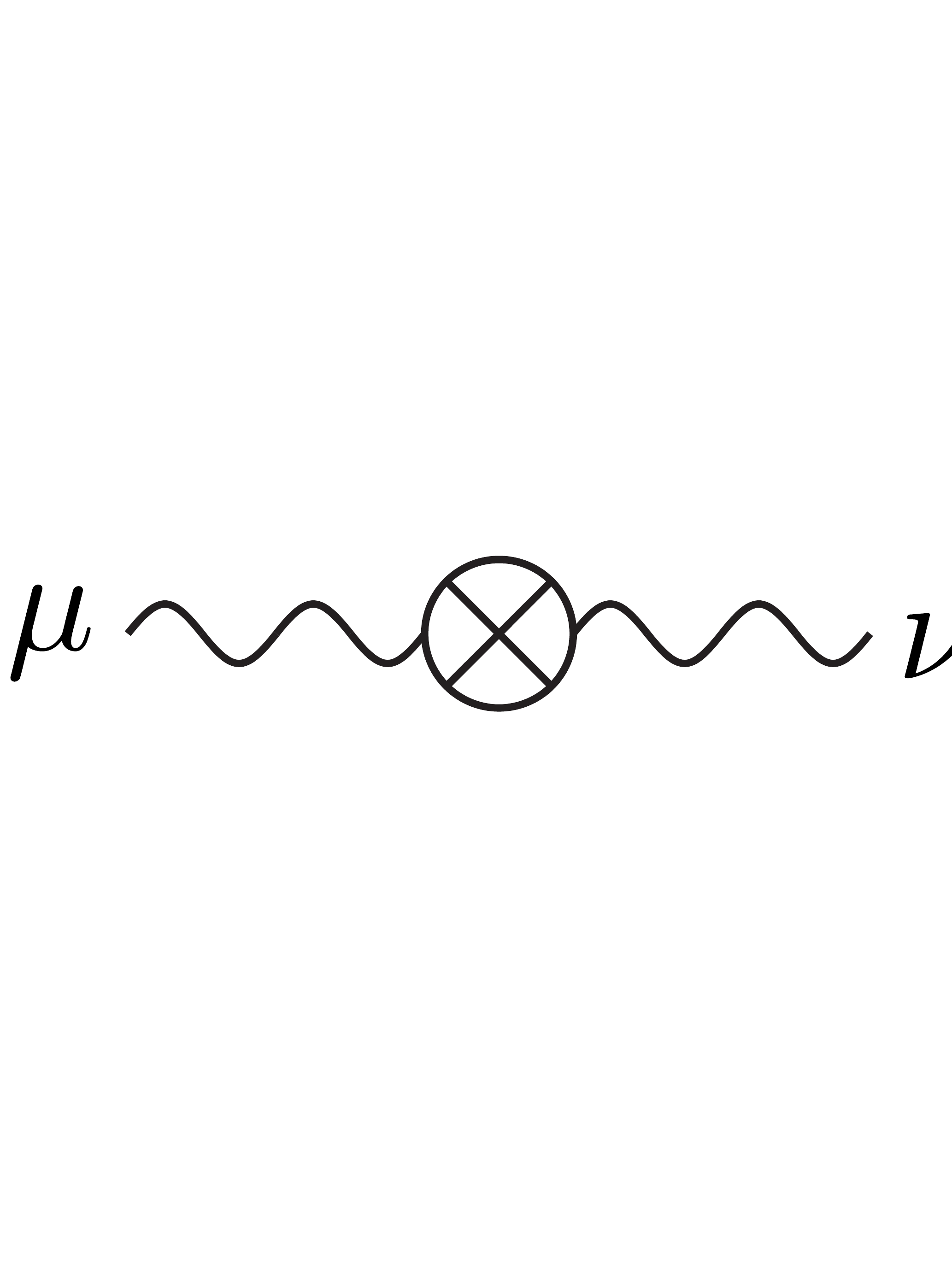}} &=& -i(g^{\mu\nu}q^2-q^\mu q^\nu)\delta_3
        \\\label{fermion}
        \raisebox{-15mm}{\includegraphics[scale=.1]{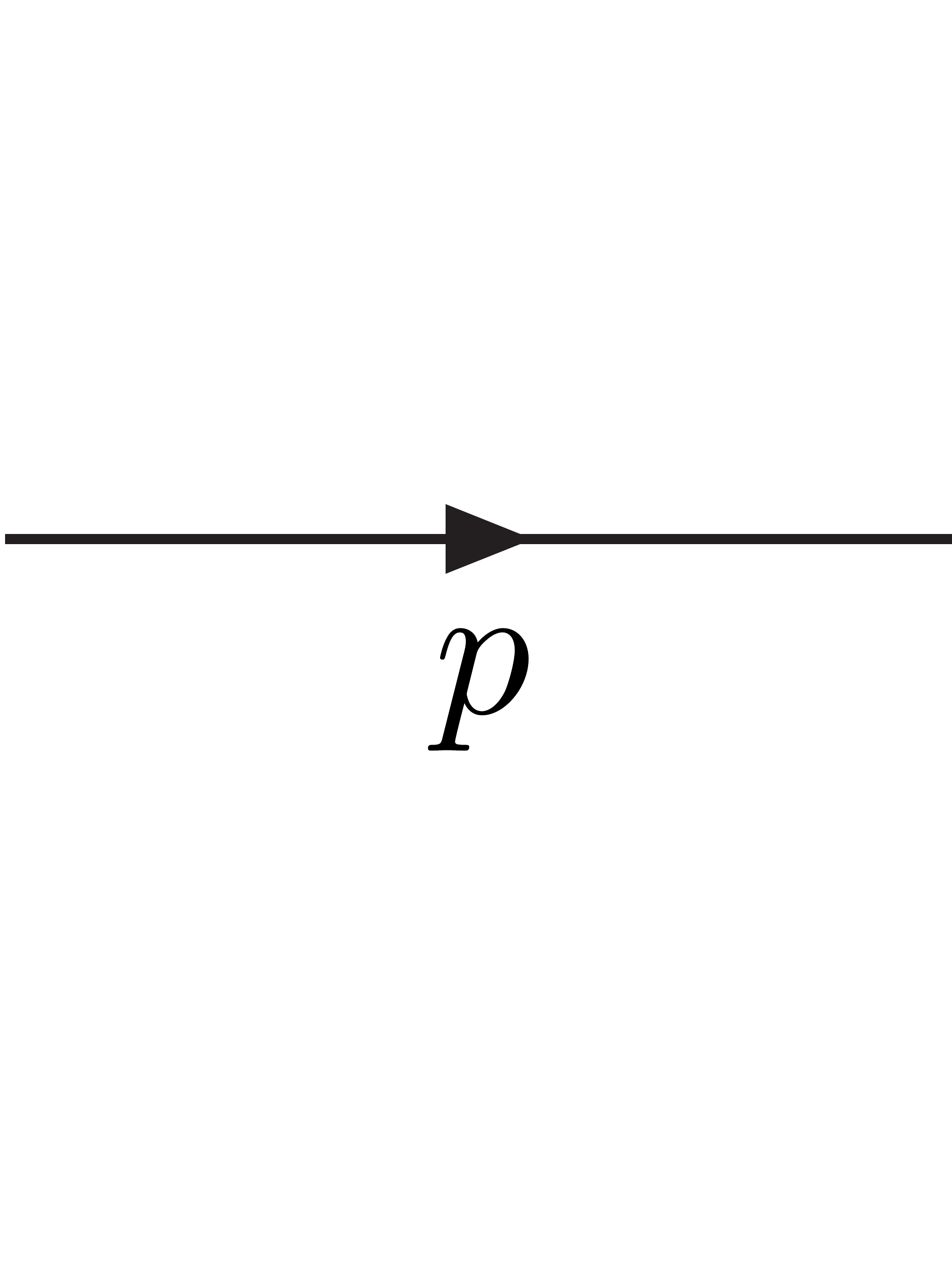}} &=& \frac{i}{p\hspace{-.23cm}\diagup-m+i\epsilon}
        \\\label{delta2m}
        \raisebox{-13mm}{\includegraphics[scale=.1]{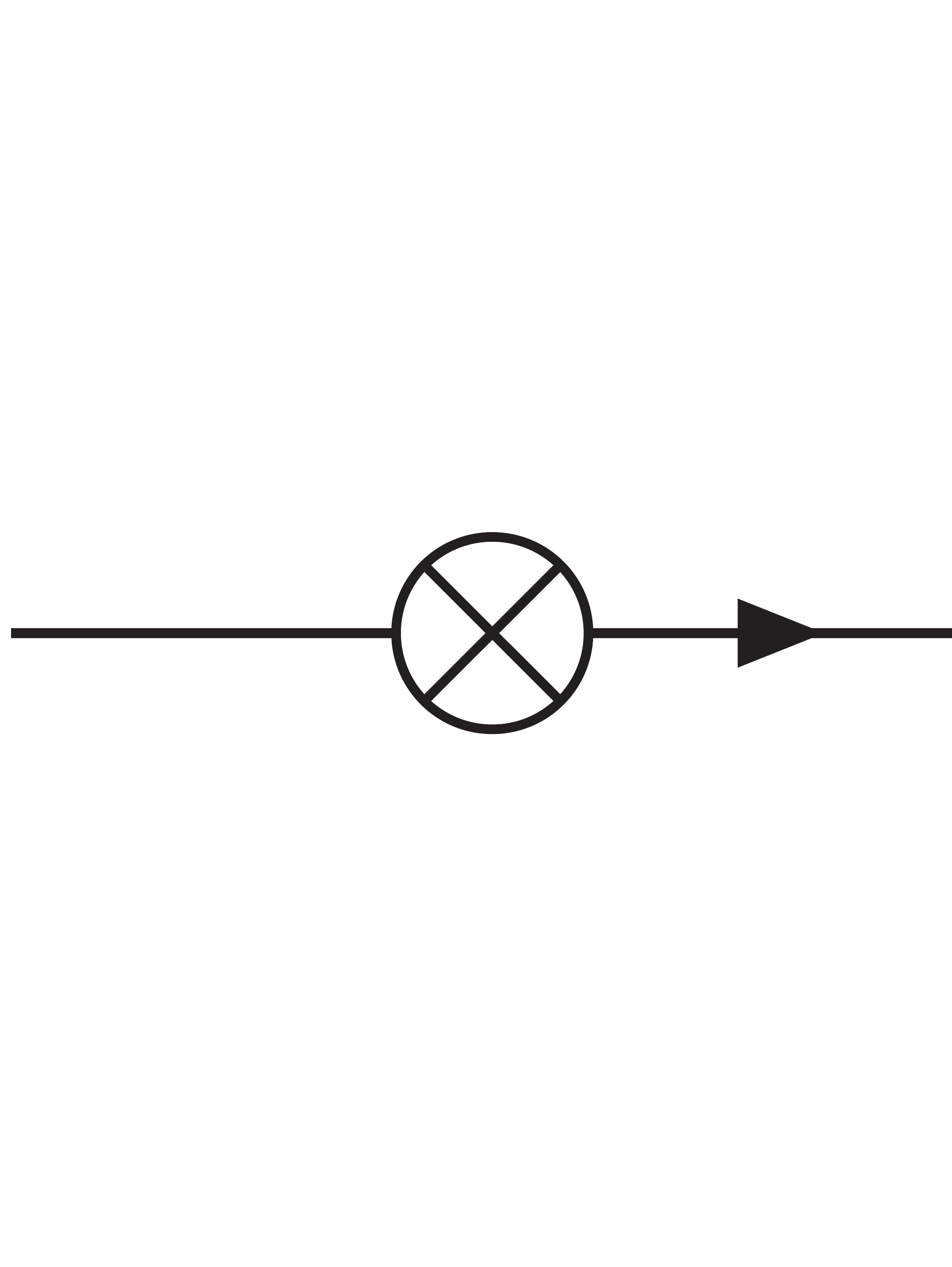}}&=& i(p\hspace{-.23cm}\diagup\delta_2-\delta_m-m\delta_2).
        \end{eqnarray}
        
        We use the following notations:
        \begin{eqnarray}\label{A.2}
        \raisebox{-13mm}{\includegraphics[scale=.1]{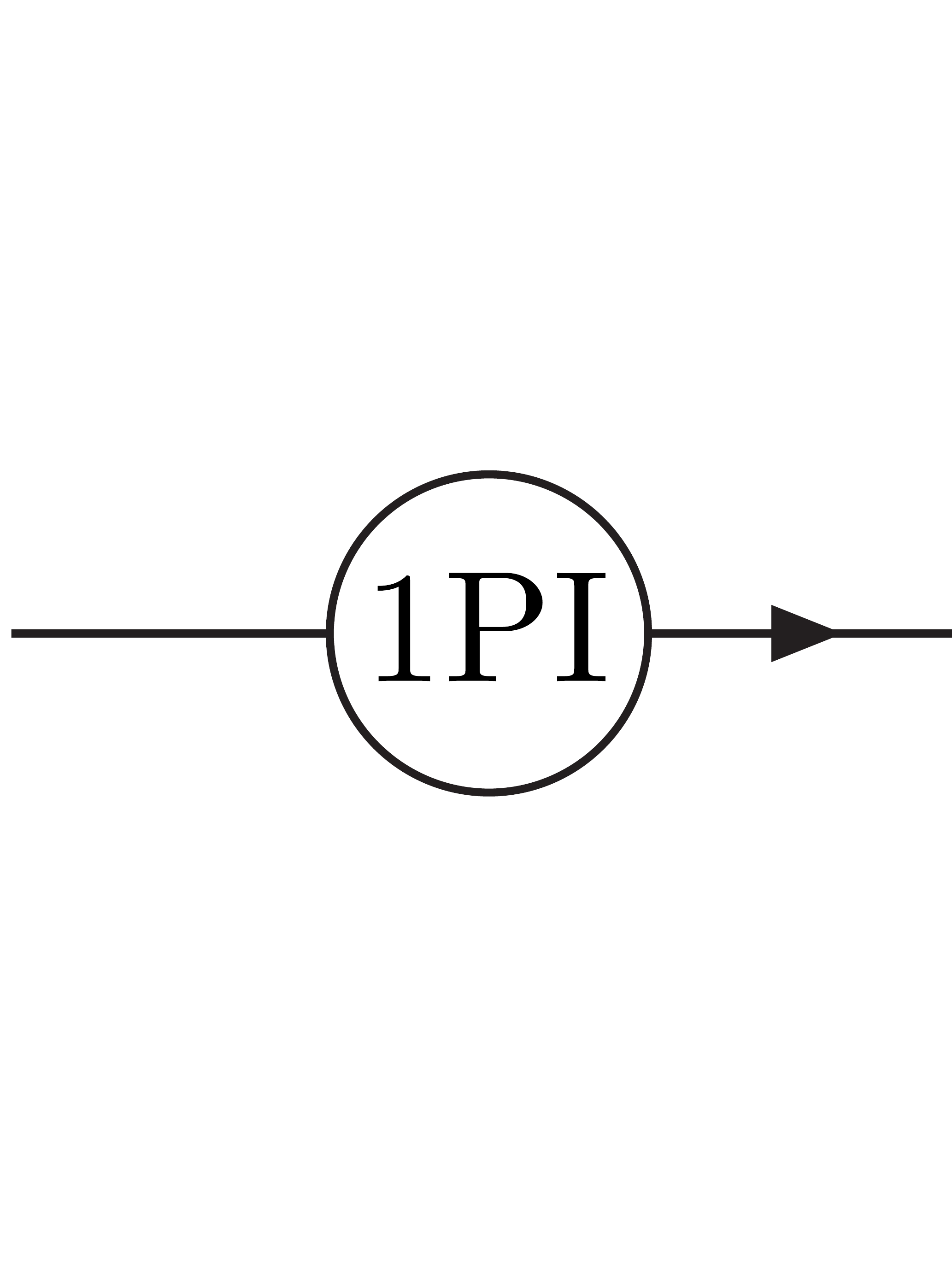}}&=&-i\Sigma(p\hspace{-.15cm}/)
        \\\label{A}
        \raisebox{-13mm}{\includegraphics[scale=.1]{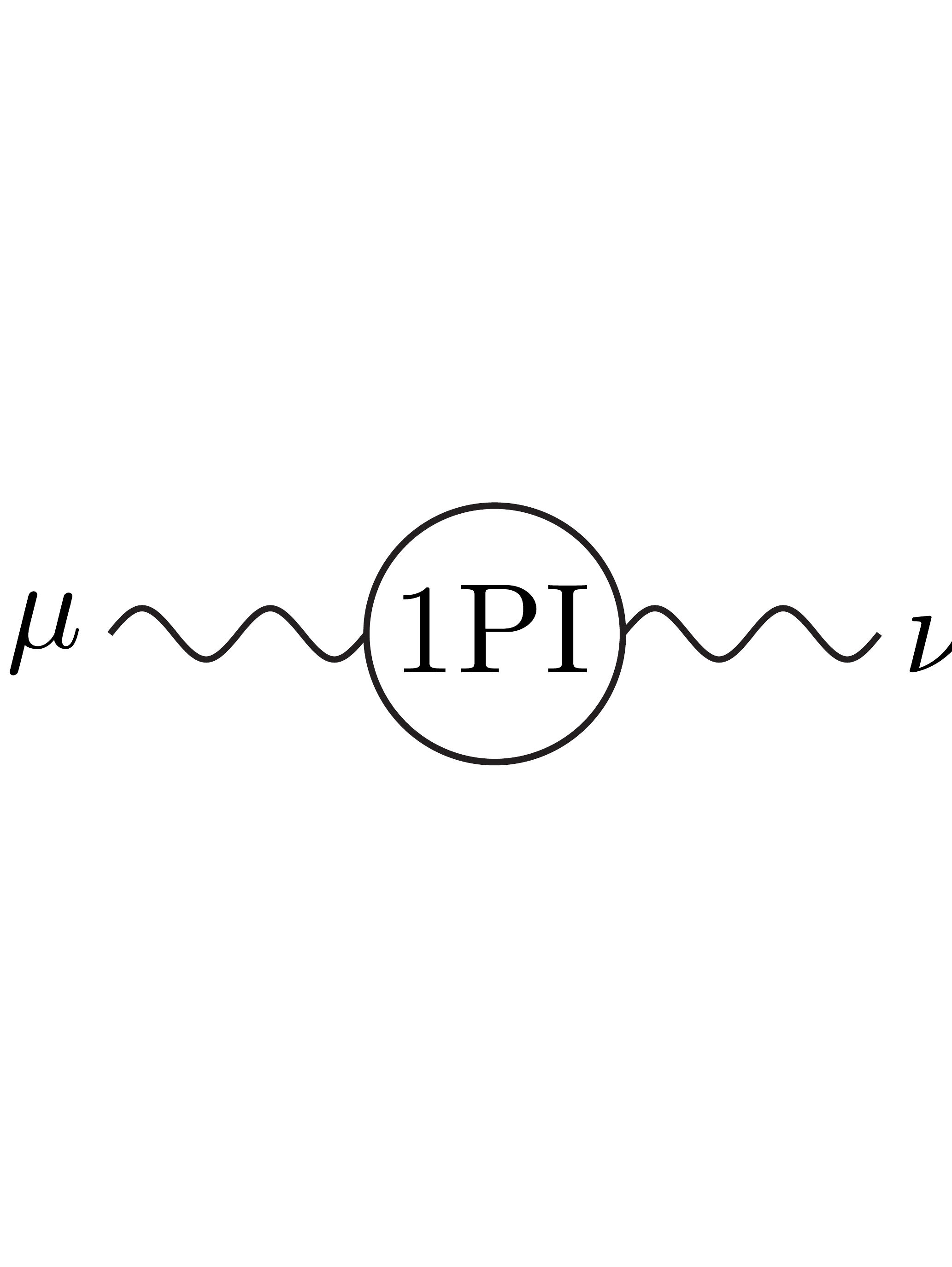}}&=&i \Pi^{\mu\nu}(q) =i(g^{\mu\nu}q^2-q^\mu q^\nu) \Pi(q^2),\\\label{A.3}
        \raisebox{-7mm}{\includegraphics[scale=.071]{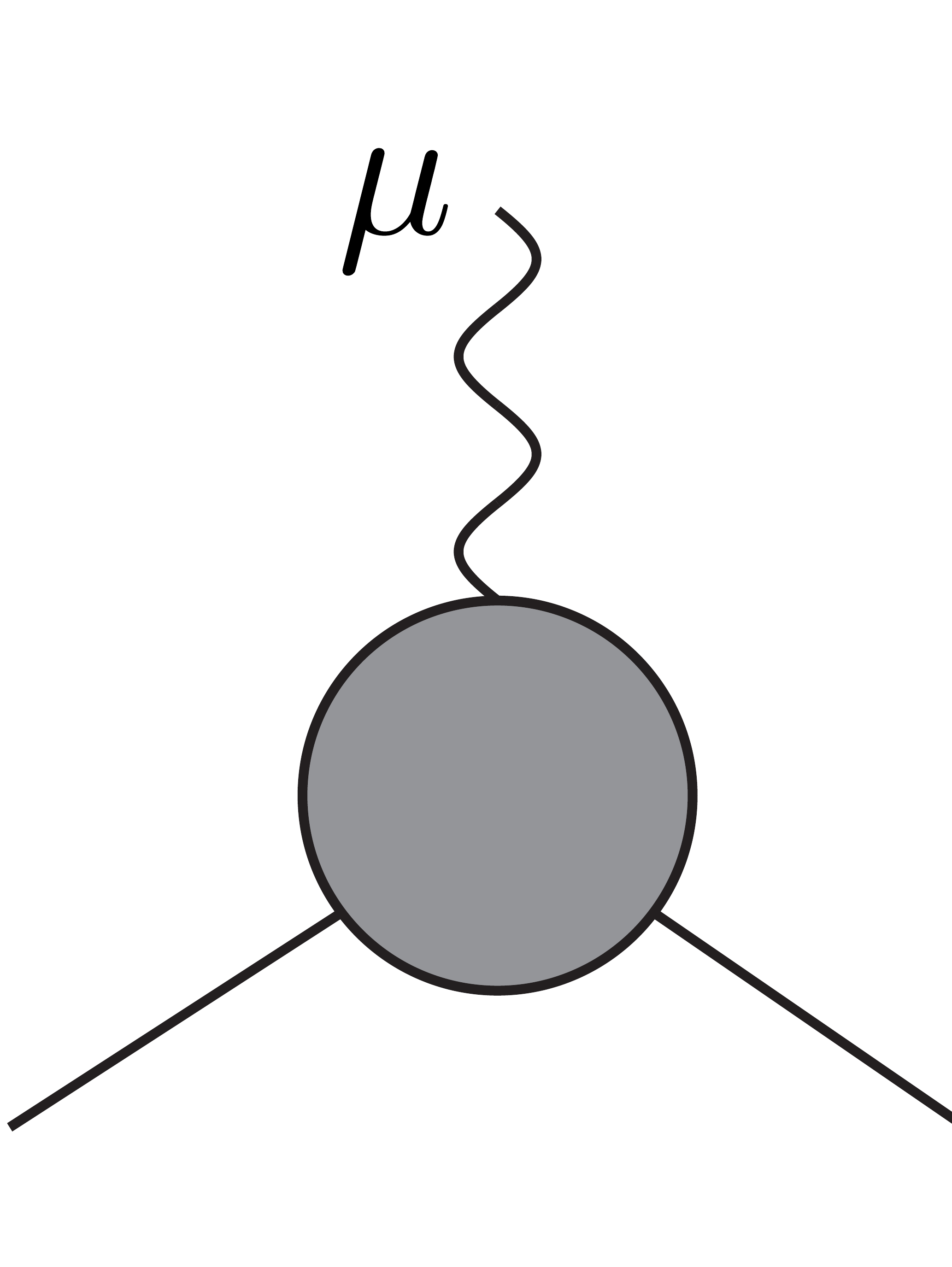}}&=&-ie\Gamma^{\mu}(p',p).
        \end{eqnarray}
        \vskip .5cm
        Here `1PI' denotes a \textit{one-particle irreducible} diagram which is the sum of any diagram that cannot split in two by removing a single line. To fix the pole of the fermion propagator at the physical mass $ m $ we need two renormalization conditions (see \cite{N}):
        \begin{eqnarray}
        \Sigma(p\hspace{-.15cm}/=m)=0\\
        \frac{d\Sigma(p{\hspace{-.15cm}/})}{dp\hspace{-.15cm}/}\bigg\vert_{p\hspace{-.13cm}{/}=m}=0.
        \end{eqnarray}
        The renormalization condition which fixes the mass of the photon to zero is
        \begin{eqnarray}
        \Pi(q^{2}=0)=0.
        \end{eqnarray}
        Given the above conditions, finally, the physical electron charge is derived by the following renormalization condition:
        \begin{eqnarray}
        -ie\Gamma^{\mu}(p'-p=0)=-ie\gamma^\mu.
        \end{eqnarray}
        
        Now, using the dimensional regularization we are able to compute $-i\Sigma(p\hspace{-.15cm}/)$, $i\Pi(q^{2})$ and $-ie\Gamma^{\mu}(p',p)$. Applying the above renormalization conditions, up to leading order in $ \alpha $, the divergent parts of the counterterms are derived as \cite{HH}
        \begin{eqnarray}
        \delta_{2}\sim-\frac{e^2}{8\pi^{2}\epsilon}\label{delta22},\\
        \delta_{m}\sim-\frac{3me^{2}}{8\pi^{2}\epsilon}\label{deltamm},\\
        \delta_{3}\sim-\frac{e^{2}}{6\pi^{2}\epsilon}\label{delta33},\\
        \delta_{1}=\delta_{2}\sim-\frac{e^{2}}{8\pi^{2}\epsilon}\label{delta11},
        \end{eqnarray}
        where $d=4-\epsilon$ is the spacetime dimension so that we should take the limit $\epsilon\to0$. As a matter of fact, these counterterms are able to remove all UV divergences of the QED theory in free space.
        \section{Renormalization in position space}\label{sec 3}

        In this section we survey the renormalization for QED in coordinate space within the renormalized perturbation theory. Naturally, when a systematic treatment of the renormalization program is done, the counterterms automatically turn out to be dependent on the functional form of the fields. In addition, the RG may lead to position dependent mass and charge, as a manifestation of the explicitly broken  translational symmetry of the system. It is worth mentioning that our main scheme is in accordance with the standard renormalization approach in momentum space where we have the translational invariance. In the next three subsections we separately consider electron self-energy, photon self-energy and vertex correction, and derive the counterterms  by imposing proper renormalization conditions in the configuration space.
        \subsubsection{Electron Self-Energy}
         According to the Lagrangian (\ref{Lagrangian}), the perturbation expansion of the full electron propagator up to order $\alpha$ is
        \begin{equation}\label{LL}
        -i\Sigma=\raisebox{-13mm}{\includegraphics[scale=.1]{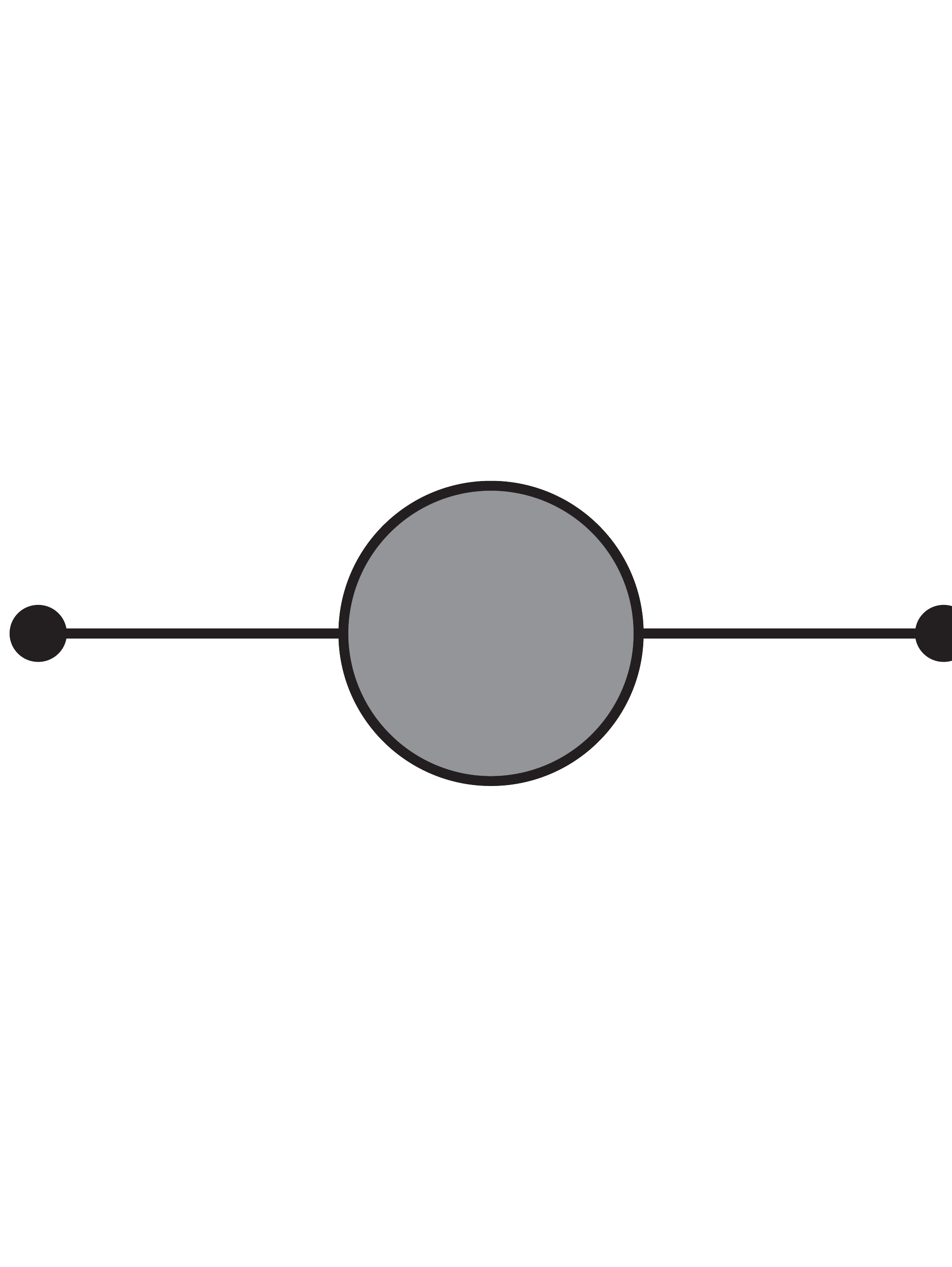}}=\raisebox{-13mm}{\includegraphics[scale=.1]{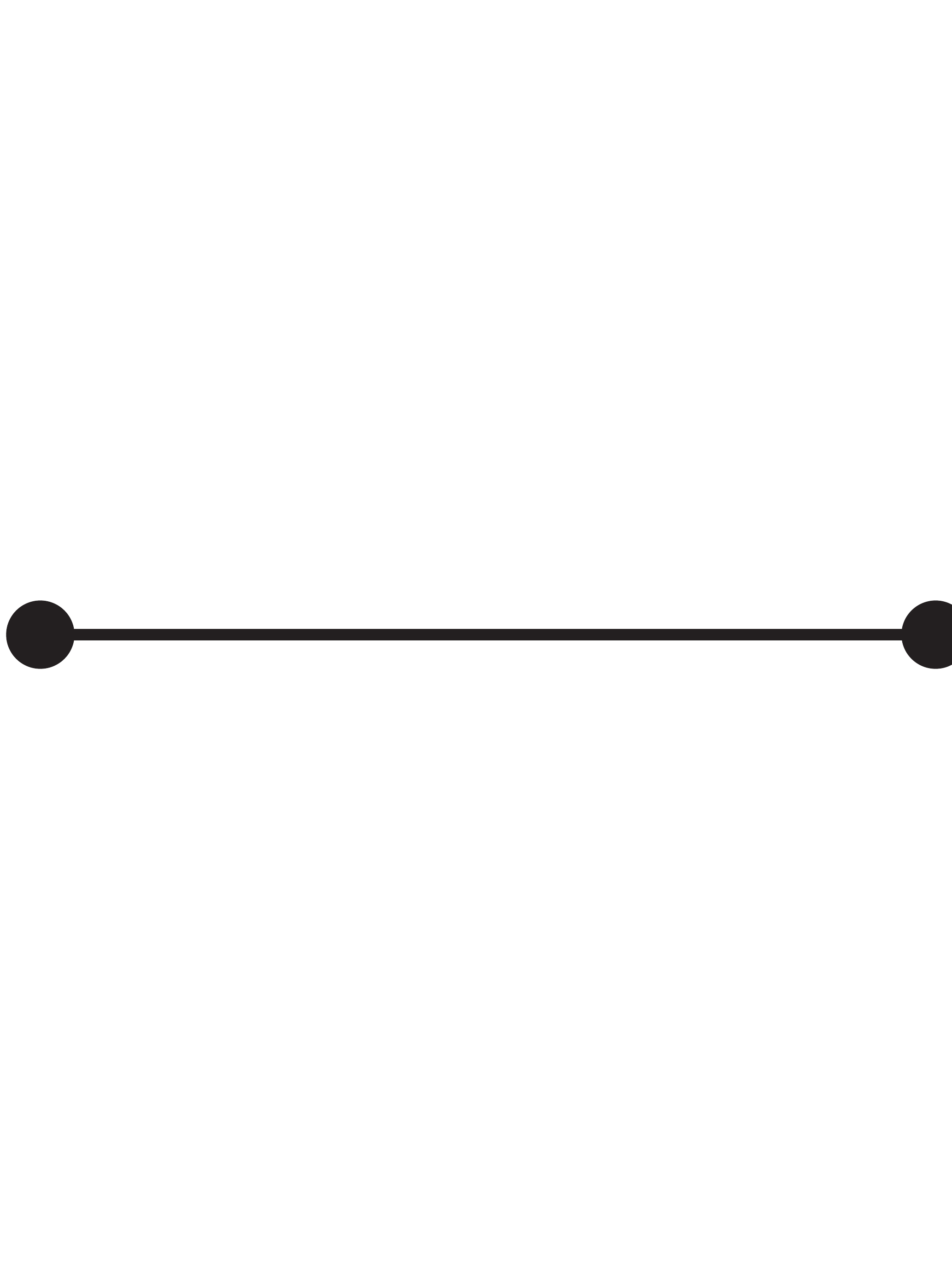}}+\raisebox{-12.2mm}{\includegraphics[scale=.1]{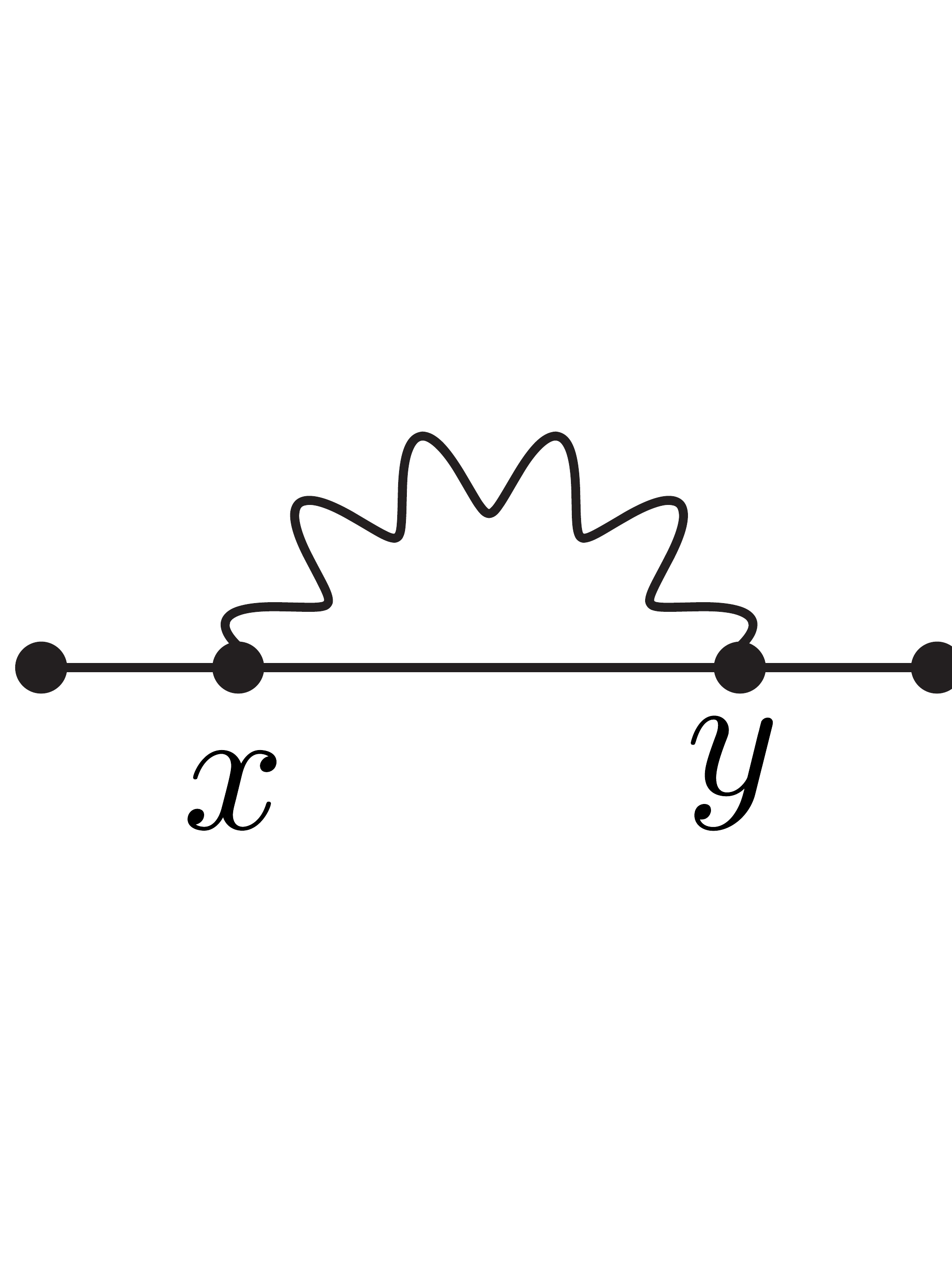}}+\raisebox{-14.4mm}{\includegraphics[scale=.1]{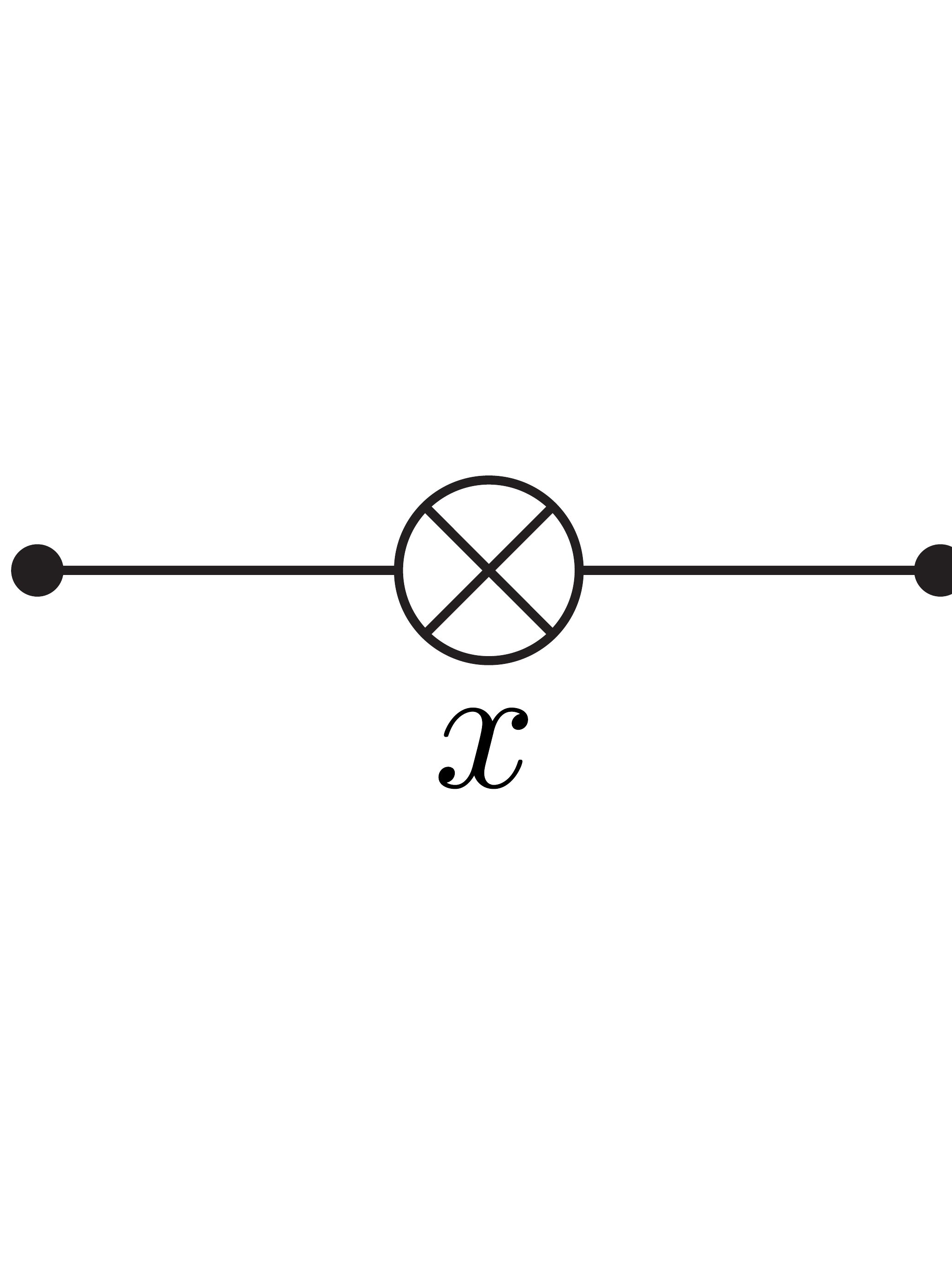}}.
        \end{equation}
        We choose our renormalization condition in such a way that pole of the first term of right hand side (RHS) gives  the physical mass $m$ at $x=x_0$. This requires that the sum of remaining diagrams, which we call it $-i\widetilde{\Sigma}(x)$ vanishes at this point, namely
        \begin{equation}\label{fig1}
        -i\widetilde{\Sigma}(x)\Bigg|_{x=x_0}=\bigg(\raisebox{-12.2mm}{\includegraphics[scale=.1]{241.pdf}}+\raisebox{-14.4mm}{\includegraphics[scale=.1]{251.pdf}}\bigg)_{x=x_0}=0,\qquad \mbox{and}\qquad \frac{d\left[-i\widetilde{\Sigma}(x)\right]}{dx}\Bigg\vert_{x=x_0}=0.
        \end{equation}
        We can write  $ - i\widetilde\Sigma$ to order  $\alpha$ as 
                \begin{eqnarray}\label{sigmatild}
        -i\widetilde{\Sigma}(x)=\int{d^d}y \overline \psi(y) \left[- i\Sigma _2(x,y) \right]\psi(x)   + \overline \psi(x)  \left[ - \delta _2(x)\partial\hspace{-.18cm}/  - im\delta _2(x)  -i \delta _m(x) \right]\psi(x)
        \end{eqnarray}
        Thus the first condition in Eq. (\ref{fig1}) yields
        \begin{eqnarray}
        -i\widetilde{\Sigma}(x_0)&=&\left\{\int{d^d}y \overline \psi(y) \left[- i\Sigma _2(x,y) \right]\psi(x)   + \overline \psi(x)  \left[ - \delta _2(x)\partial\hspace{-.18cm}/  - im\delta _2(x)  -i \delta _m(x) \right]\psi(x) \right\}_{x=x_0}\\&=& 0,
        \end{eqnarray}
        where $ - i\Sigma _2$ is $O(\alpha)$ electron self-energy diagram.
        Now, using Dirac equation $ \left( {i\partial\hspace{-.18cm}/  - m} \right)\psi  = 0$, up to order $\alpha$ we obtain
        
        \begin{equation}\label{deltam}
        {\delta _m} =\frac{- 1 }{\overline \psi(x_0)  \psi(x_0) }\int {d^d}y  \overline \psi(y)   {{\Sigma _2(x,y)}}\psi(x)\bigg|_{x=x_0}.
        \end{equation}
        To simplify the second condition in Eq. (\ref{fig1}) we note that the $\widetilde{\Sigma}(x)$ is, in fact, a function of $\overline {\psi}(x), \psi(x), \partial\hspace{-.18cm}{/} \overline{\psi}(x) $ and $\partial\hspace{-.18cm}{/} \psi(x)$ so that
        \begin{eqnarray}\label{der.}
        \frac{d\widetilde{\Sigma}(x)}{dx}=\frac{\partial\psi}{\partial x}\frac{\partial\widetilde{\Sigma}}{\partial\psi}+\frac{\partial \overline\psi}{\partial x}\frac{\partial\widetilde{\Sigma}}{\partial\overline\psi}+\frac{\partial( \partial\hspace{-.18cm}{/} \overline\psi)}{\partial x}\frac{\partial\widetilde{\Sigma}}{\partial(\partial\hspace{-.18cm}{/}\overline\psi)}+\frac{\partial(\partial\hspace{-.18cm}{/} \psi)}{\partial x}\frac{\partial\widetilde{\Sigma}}{\partial(\partial\hspace{-.18cm}{/}\psi)}.
        \end{eqnarray}
        Due to the opposite sign of the momentum for particles and anti-particles the first two terms in  cancel each other. The third term is also zero, because there is no derivative of $\overline{\psi}$ in  $\widetilde{\Sigma}$ . Thus, we obtain
        \begin{equation}\label{con}
        \frac{{\partial\left[ { - i\widetilde\Sigma(x) } \right]}}{{\partial\left({\partial\hspace{-.18cm}{/} \psi } \right)}}\Bigg\vert_{x=x_0} = 0.
        \end{equation}
        We can derive $\delta_2(x_0)$ by using the above equation and Eq. (\ref{sigmatild})
        \begin{eqnarray}
        \frac{{\partial\left[ { - i\widetilde\Sigma(x) } \right]}}{{\partial\left({\partial\hspace{-.18cm}{/} \psi } \right)}}\Bigg\vert_{x=x_0}&=& \int d^dy \frac{\partial\left[\overline\psi(y)(- i\Sigma_2(x,y) )\psi(x)\right]}{\partial\left(\partial\hspace{-.18cm}/ \psi(x)\right)}\Bigg\vert_{x=x_0} - \overline\psi(x_0) \delta_2(x_0)= 0\\\Rightarrow \delta_2& =&\frac{1}{\overline\psi(x_0)}\int d^dy \frac{\partial\left[\overline\psi(y)( - i\Sigma_2(x,y)) \psi(x)\right]}{ \partial \left(\partial\hspace{-.18cm}/ \psi(x)\right)}\bigg\vert_{x=x_0}. \label{delta2}
        \end{eqnarray}

        \subsubsection{Photon Self-Energy}
        For the photon propagator we again expand the full propagator as\vspace{-1cm}
        \begin{equation}
        i\Pi=\raisebox{-13mm}{\includegraphics[scale=.1]{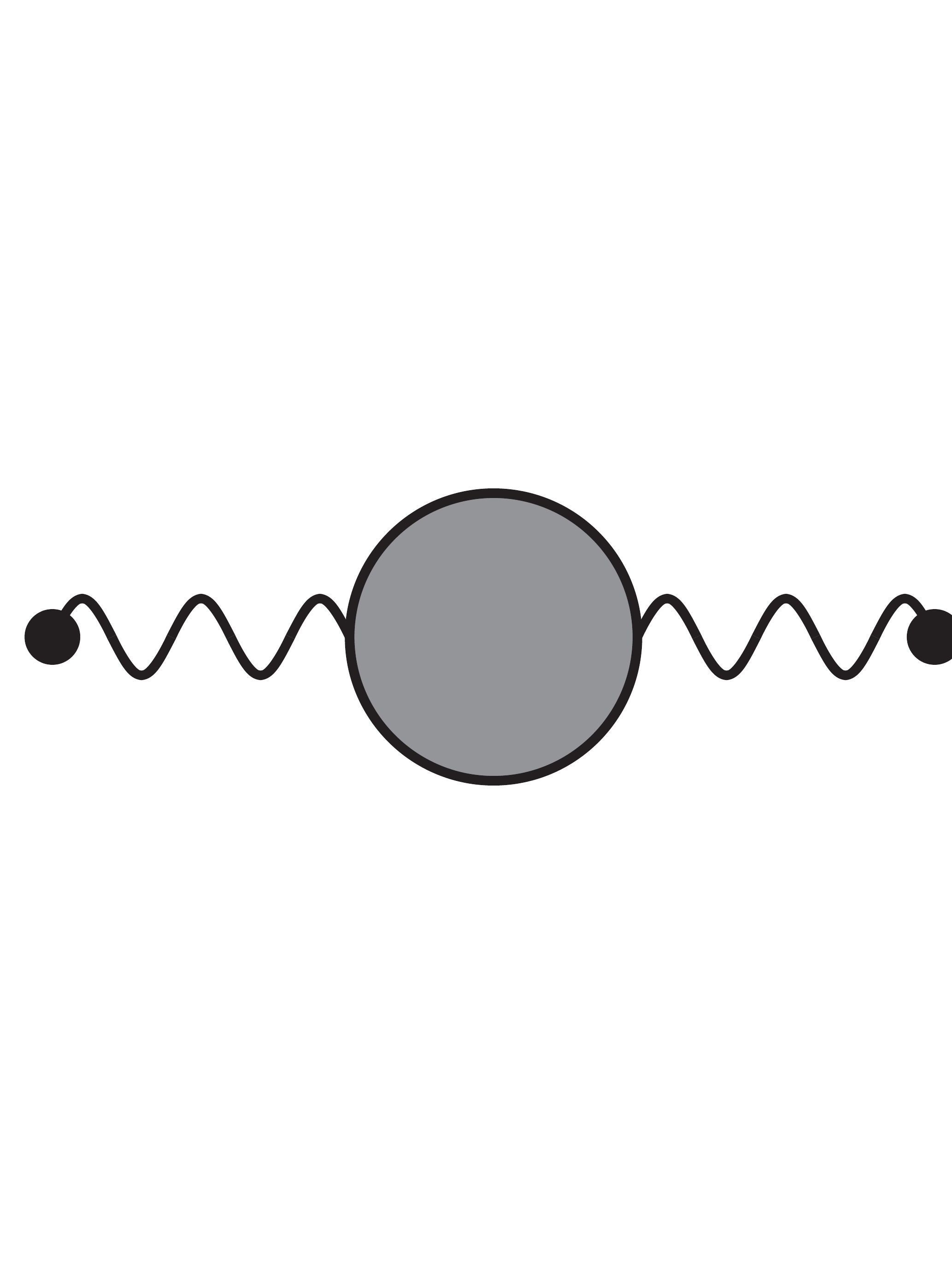}}=\raisebox{-10.3mm}{\includegraphics[scale=.08]{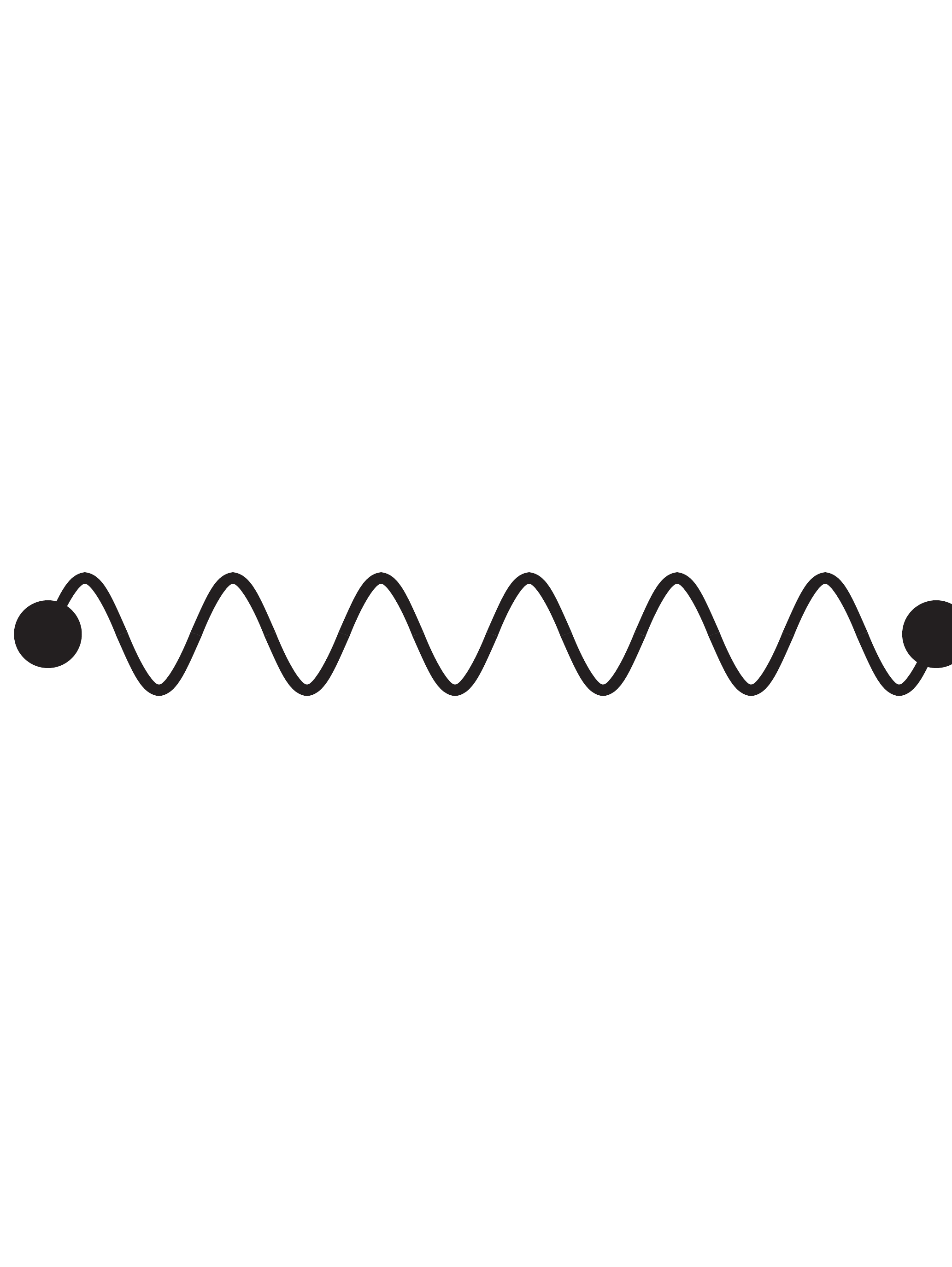}}+\raisebox{-13.2mm}{\includegraphics[scale=.1]{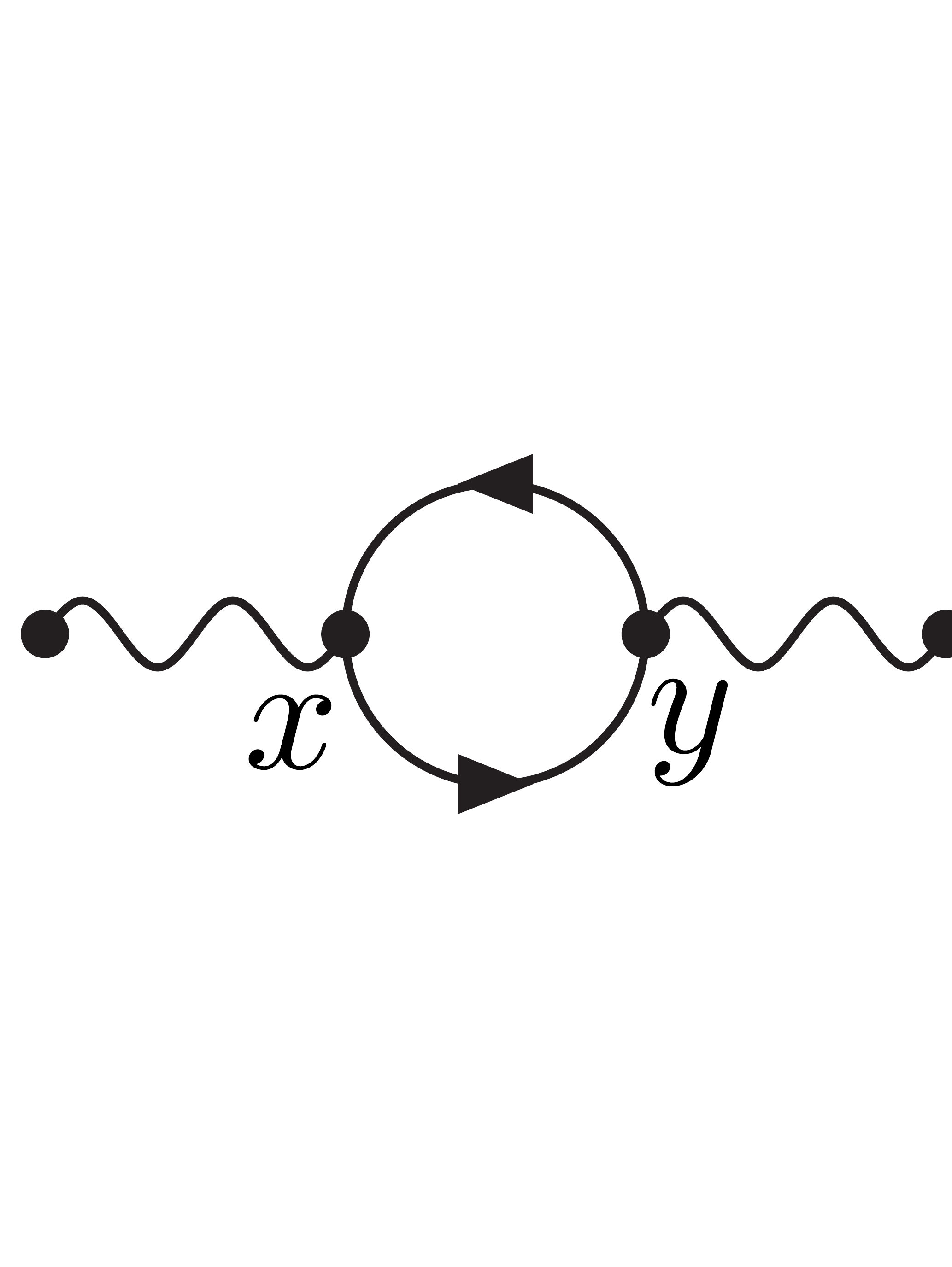}}+\raisebox{-14.4mm}{\includegraphics[scale=.1]{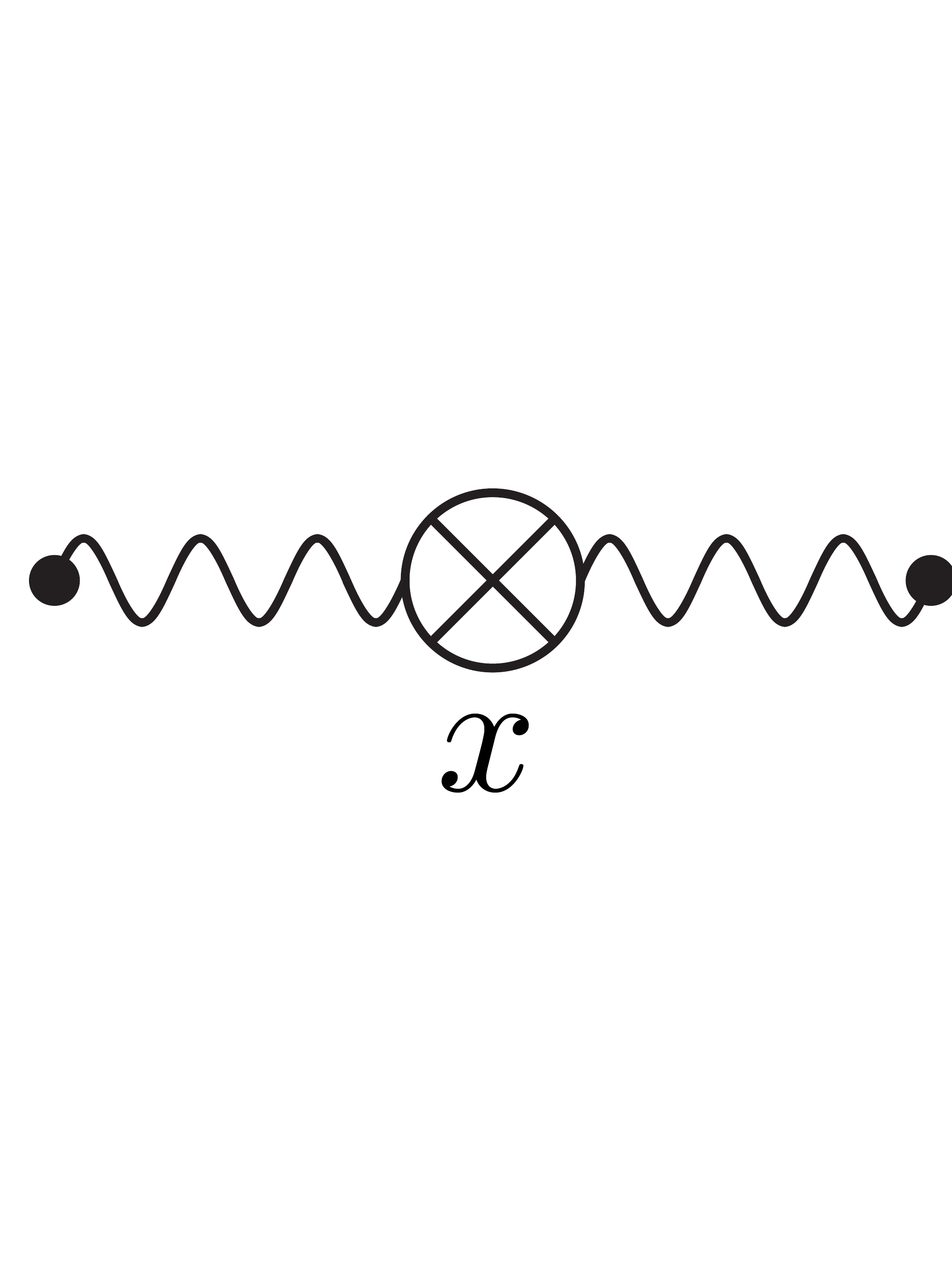}}+\dots.
        \end{equation}
        To have a massless photon, at $x=x_0$, we need only the first term on the RHS with a pole which definitely fixed on zero. Therefore, the rest of the perturbation series must vanish, so that up to order $\alpha$ we have
        \begin{equation}\label{fig2}
        i\widetilde{\Pi}(x)\bigg|_{x=x_0}=\bigg(\raisebox{-13.2mm}{\includegraphics[scale=.1]{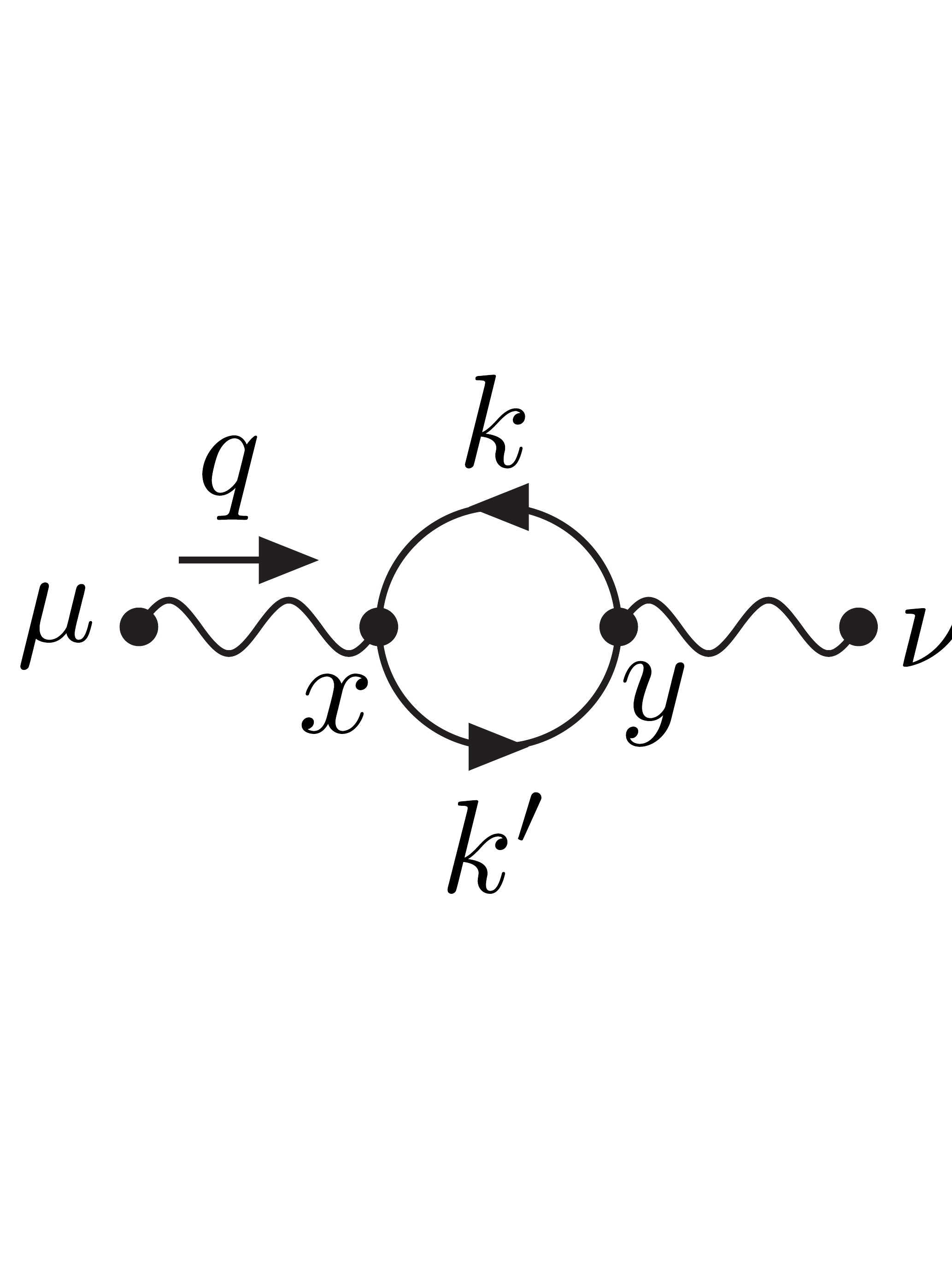}}+\raisebox{-14.4mm}{\includegraphics[scale=.1]{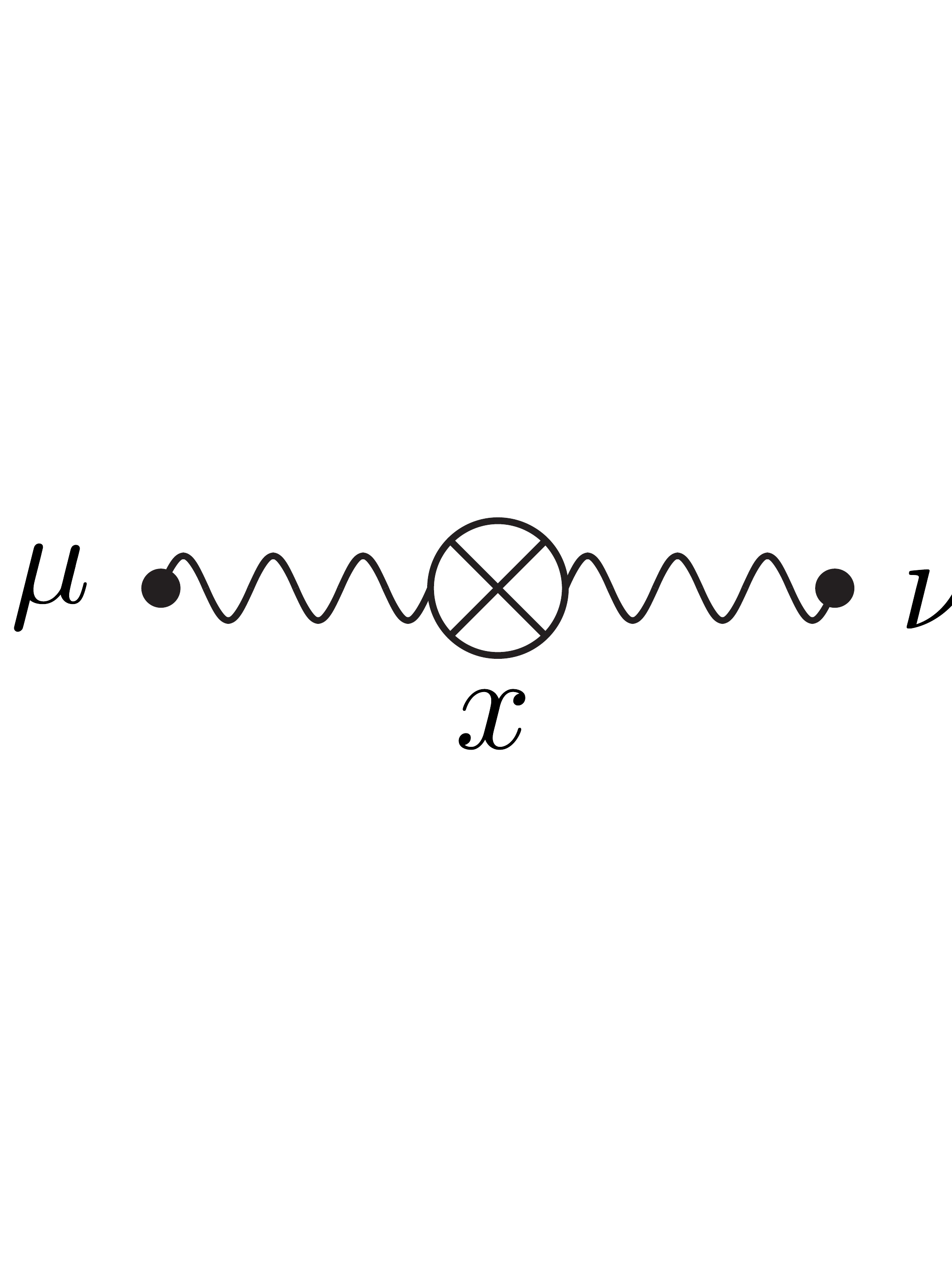}}\bigg)_{x=x_0}=0
        \end{equation}
        or equivalently,
        \begin{equation}
        i\widetilde{\Pi}(x_0)=\left\{\int {d^d}y{\widetilde{A}^{ *}_\mu(y)}\left[i{\Pi _2^{\mu \nu }}(x,y)\right]{\widetilde{A}_\nu }\left(x \right) + \widetilde{A}^{ * }_\mu\left( x \right)\delta _3(x)\left[ - i \left(g^{\mu \nu}\partial ^2- \partial^\mu \partial^\nu \right)\right]\widetilde{A}_\nu \left( x \right)\right\}_{x=x_0}=0,
        \end{equation}
        where $i\Pi_2 ^{\mu \nu}(x,y)$ is $O(\alpha)$ photon self-energy diagram.
        Therefore,
        \begin{equation}\label{delta3}
        \delta _3 = \int d^d y\frac{-\widetilde{A}^{ * }_\mu(y) \Pi_2 ^{\mu \nu }(x,y)\widetilde{A}_\nu \left( x \right)}{\widetilde{A}^*_\mu(x)(g^{\mu \nu }\partial ^2-\partial^\mu \partial^\nu)\widetilde{A}_\nu \left( x \right)}\Bigg\vert_{x=x_0}.
        \end{equation}
        
        \subsubsection{Vertex Correction}
        Formally, the vertex corrections give us the physical charge of electron. Diagrammatically we have
        
        \begin{equation}\label{FFF}
        -ie{\Gamma}^\mu(x)=\raisebox{-8mm}{\includegraphics[scale=.08]{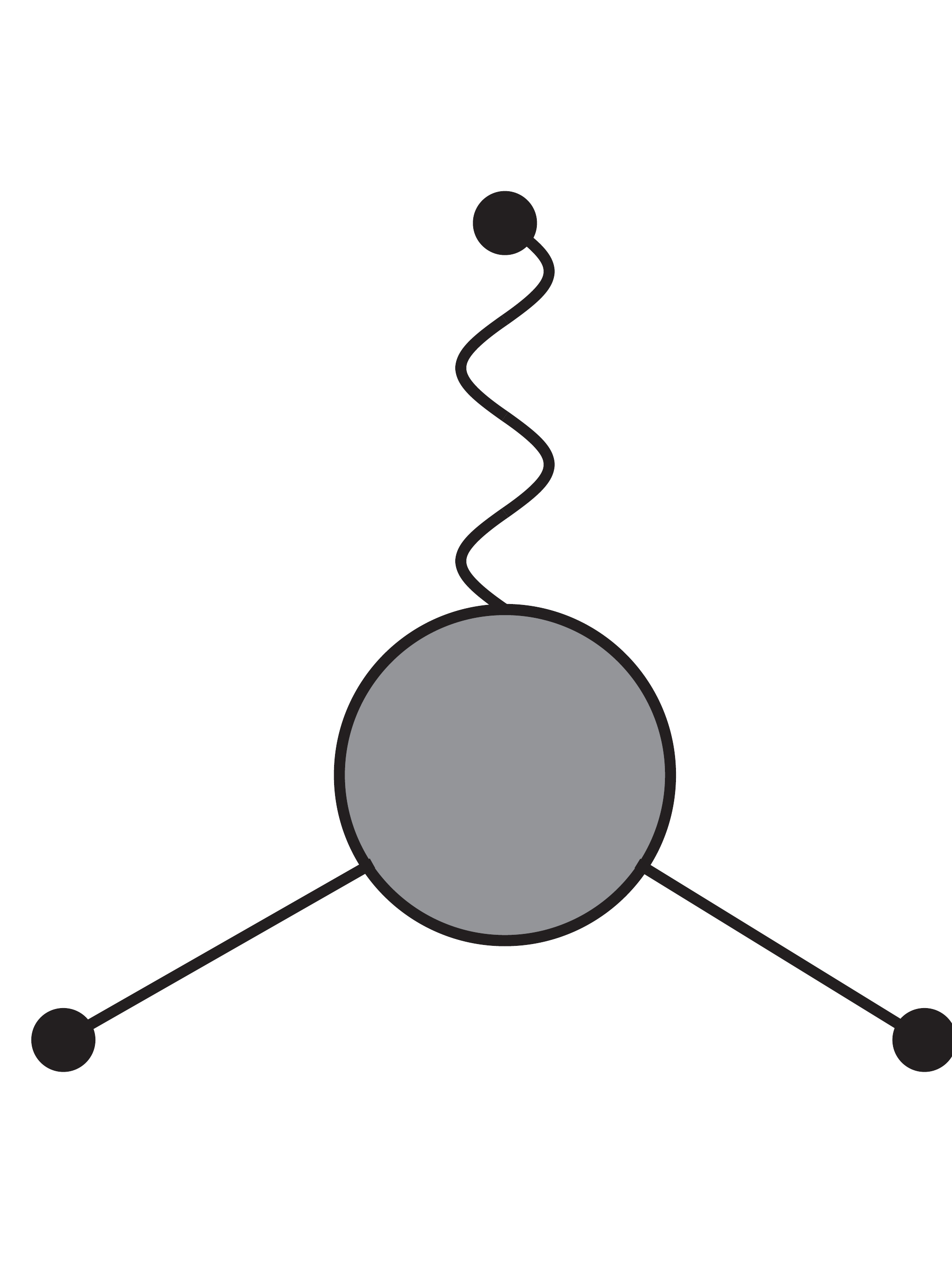}}=\raisebox{-8.3mm}{\includegraphics[scale=.08]{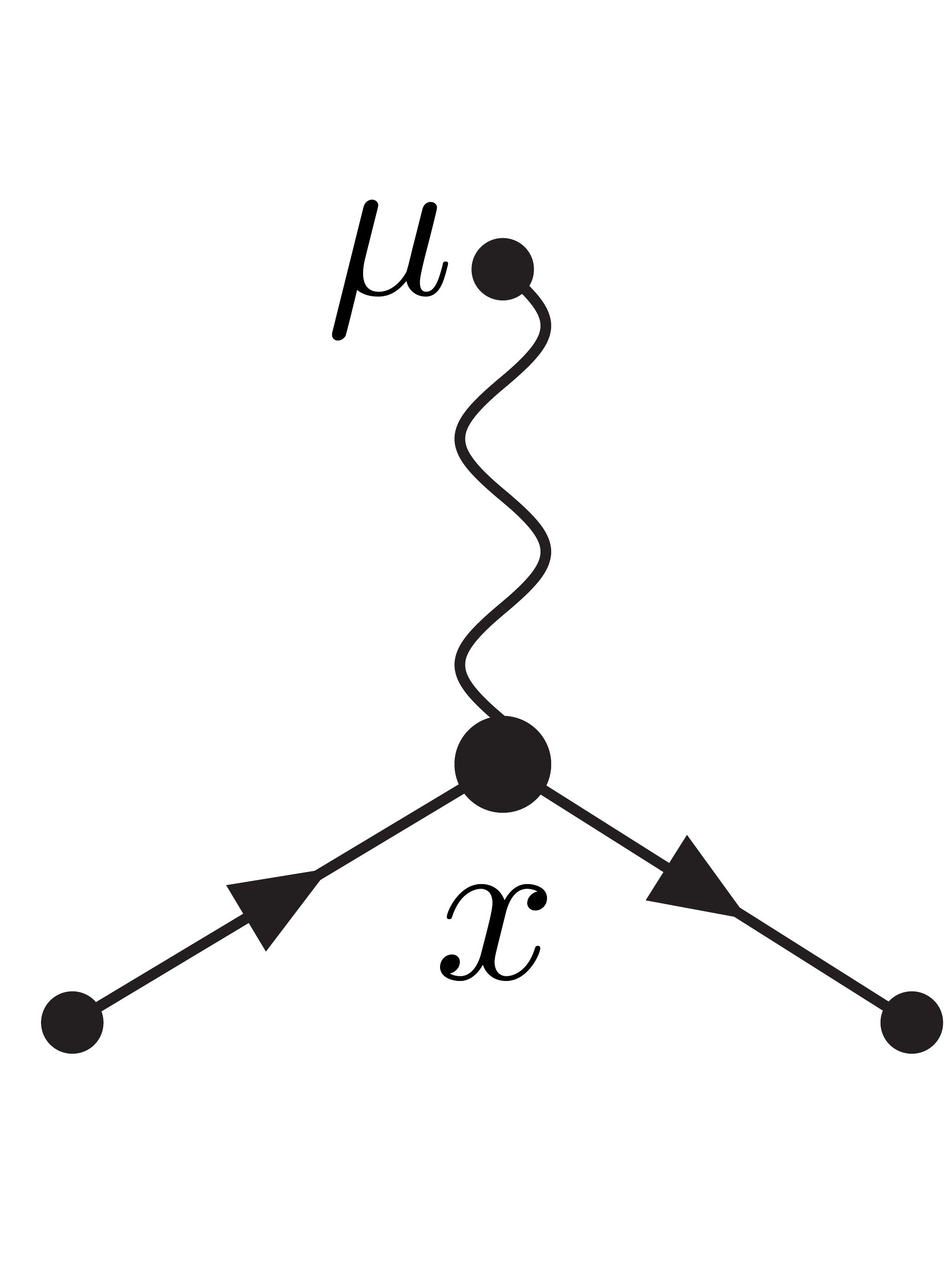}}+\raisebox{-10mm}{\includegraphics[scale=.09]{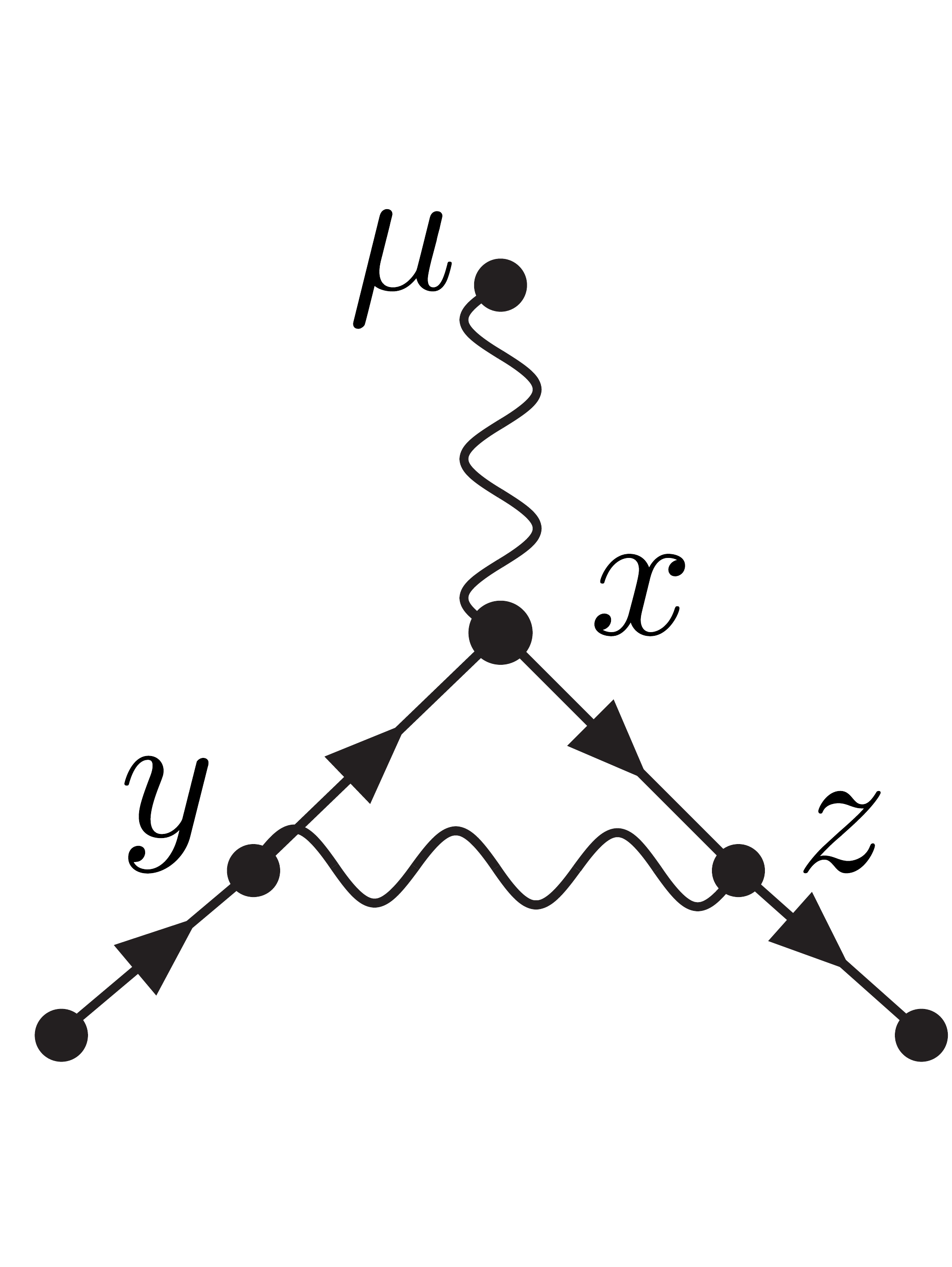}}+\raisebox{-8mm}{\includegraphics[scale=.075]{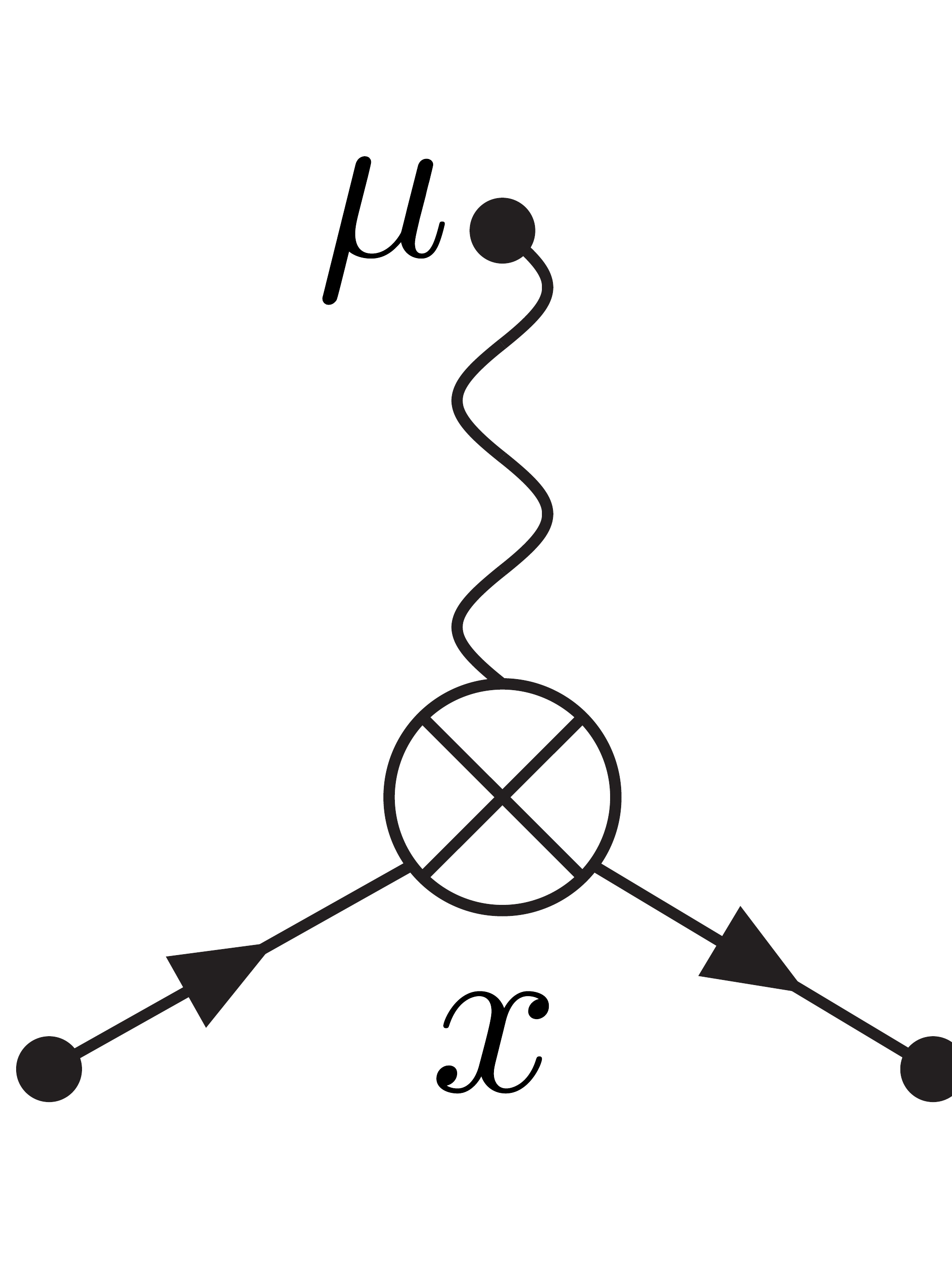}}+\dots
        \end{equation}
        Our renormalization condition for the electron charge is to fix it on physical $e$ at $x=x_0$. We can do this by using the first term on RHS of Eq. (\ref{FFF}), so that the remaining diagrams should cancel each other,
        
        \begin{equation}\label{vertex1}
        -ie\widetilde{\Gamma}^\mu(x_0)=\myl(\raisebox{-10mm}{\includegraphics[scale=.09]{341.pdf}}+\raisebox{-8mm}{\includegraphics[scale=.075]{351.pdf}}\myr)_{x=x_0}.
        \end{equation}
        We can equivalently write the above equation as,
        \begin{equation}
        -ie\widetilde{\Gamma}^\mu(x_0)=\left\{\int d^dy\, d^dz\,\overline \psi(z)  [-ie \delta\Gamma^\mu(x,y,z)]\psi(y) +  \overline \psi(x)  \left[  -ie\delta_1(x)\gamma ^\mu \right]\psi(x) \right\}_{x=x_0}= 0,
        \end{equation}
        where $ -ie\delta\Gamma^\rho$ is the vertex correction diagram to order $\alpha$. Therefore we find
        \begin{equation}\label{delta1}
        \delta_1\gamma ^{\mu}=\int {d^d}y \, {d^d}z\frac{\overline \psi(z) \delta\Gamma^\mu(x,y,z) \psi(y) }{ \overline \psi(x) \psi(x)}\Bigg|_{x=x_0}.
        \end{equation}
       Accordingly we may derive counterterms required for renormalzation of QED in coordinate space. These counterterms could be applied for problems in which the translational invariance breaks explicitly. Obviously if we work in free space, with the translational symmetry, they should reduce to those in the standard prevalent derived in momentum space. We show this equivalence in the next section.
        
        \section{Comparison to Momentum Space (Free Space)}\label {sec 4}
        In this section, as a special case, we compare our results with renormalization of QED in free space. In free space, the wave functions of fermions and photons are considered as  plane waves. We start with the Eq. (\ref{deltam}) by inserting $\psi \left( {x} \right) = u^s(p) {e^{ - ip.x}}$ (from here on we drop the superscript $s$ for simplicity). Then, the numerator of the integrand becomes
        \begin{eqnarray}\nonumber
        \int d^d y \ \overline \psi(y)  [ { - i{\Sigma _2}(x,y)} ]\psi(x)&=&-{e^2} {  \int {{d^d}y}\,  \overline u \left( p \right){e^{  ip.y}} {\gamma ^\mu }S\left( {x - y} \right){\gamma ^\nu }D_{\mu\nu}\left( {y - x} \right)} u \left( p \right){e^{-ip.x}}\\
        &=&  -{e^2}\overline u \left( p \right)\left[ {\int {{d^d}y {\frac{{{d^d}k}}{{{{\left( {2\pi } \right)}^d}}} {\frac{{{d^d}k'}}{{{{\left( {2\pi } \right)}^d}}}{\gamma ^\mu }\frac{ k\hspace{-.18cm}/ - m}{{ k^2 - m^2}}\frac{{ \gamma _\mu }}{{{{k'}^2}}}{e^{ - i\left( {k + k' - p} \right).y}}{e^{ - i\left( {p - k' - k} \right).x}}} } } } \right]u\left( p\right),\label{azx}
        \end{eqnarray}
        where $S\left( {x - y} \right)$ and $D^{\mu\nu}\left( {y - x} \right)$ are the  propagators of fermion and photon in $d$ spacetime dimensions, respectively,
        \begin{equation}
        S\left( {x - y} \right)=\int {\frac{{{d^d}k}}{{{{\left( {2\pi } \right)}^d}}}}\frac{{{i}}}{k\hspace{-.18cm}/ -m}e^{ - ik.\left( {x - y} \right)},
        \end{equation}
        and,
        \begin{equation}
        D^{\mu\nu}\left({x - y} \right)=\int {\frac{{{d^d}k}}{{{{\left( {2\pi } \right)}^d}}}}\frac{{{-ig^{\mu\nu}}}}{k^2}e^{ - ik.\left( {x - y} \right)}.
        \end{equation}
        Integrating over position and then $k'$ in Eq. (\ref{azx}) yields
        \begin{equation}
        \int d^d y \ \overline \psi(y)  [ { - i{\Sigma _2}(x,y)} ]\psi(x)=-{e^2} \overline u \left( p \right)\left[ { \int {\frac{{{d^d}k}}{{{{\left( {2\pi } \right)}^d}}}{\gamma ^\mu }\frac{1}{{k\hspace{-.18cm}/ - m}}{\gamma _\mu }\frac{{ 1}}{{{{\left( {p - k} \right)}^2}}}} } \right]u\left( p \right).
        \end{equation}
        In terms of $\epsilon=4-d$, the above equation becomes
        \begin{eqnarray}\label{ttt}
        \int d^d y \ \overline \psi(y)[{ - i{\Sigma _2}(x,y)} ]\psi(x)&=&\overline u \left( p \right) {\frac{{ -i{e^2}}}{{8{\pi ^2}\epsilon }}\left( { - p\hspace*{-.15cm}/ + 4m} \right)} u \left( p \right)+O(\epsilon^0)\\
        &=& {\frac{{ -3im{e^2}}}{{8{\pi ^2}\epsilon }}} \overline u(p) u(p)+O(\epsilon^0).
        \end{eqnarray}
        Finally, using Eq. (\ref{deltam}) we have
        \begin{eqnarray}
        {\delta _m} &=& \frac{-1}{\overline u u }{\frac{{  3m{e^2}\overline u u}}{{8{\pi ^2}\epsilon }}} +O(\epsilon^0)\nonumber\\&=& \frac{{ - 3m{e^2}}}{{8{\pi ^2}\epsilon }}+O(\epsilon^0),
        \end{eqnarray}
        The above result is clearly independent of $x_0$, manifesting the translational invariance of the system. It is also in agreement with  Eq. (\ref{deltamm}), the standard common counterterm derived directly in free space.\\
        We similarly derive the second counterterm, $\delta_2$. Now, using Eq. (\ref{ttt}) and the fact that $\partial\hspace{-.18cm} / \psi=\partial\hspace{-.18cm}/[u(p)e^{-ip.x}]=-ip\hspace{-.15cm}/\psi$,  we can rewrite the Eq.(\ref{delta2}) as follows:
        \begin{equation}
        {\delta _2} = \frac{1}{\overline u(p)}\frac{{\partial\left[\overline u(p) {\frac{{ - i{e^2}}}{{8{\pi ^2}\epsilon }}\left( { -p\hspace{-.14cm}/ + 4m} \right)u(p) } \right]}}{{\partial\left[ {-ip\hspace{-.14cm}/u(p) } \right]}} = \frac{{ - {e^2}}}{{8{\pi ^2}\epsilon }}+O(\epsilon^0),
        \end{equation}
        which is precisely in agreement with Eq. (\ref{delta22}). Again we see that the position dependence cancels out as expected.
        
        To compute $\delta_3$ in free space, we use $\widetilde{A}_\mu(p,x)=\varepsilon^s_\mu(p) e^{-ip.x}$ in Eq. (\ref{delta3}). The numerator becomes 
        \begin{eqnarray}\nonumber
        \int d ^dy \widetilde{A}^ * _\mu(y)\left( i\Pi_2 ^{\mu \nu}\right)\widetilde{A}_\nu(x) &=& \varepsilon ^* _\mu\left[  -i e^2\int d^dy \,\gamma ^\mu  S\left( x - y \right)\gamma ^\nu S\left( y - x \right) e^{ - iq.x}e^{iq.y}\right]{\varepsilon _\nu }
        \\
        &=&\varepsilon ^*_ \mu\left[i e^2 \int d^d y \int\frac{d^dk}{{\left( 2\pi  \right)}^d} \frac{d^dk'}{{\left( {2\pi } \right)}^d}\gamma ^\mu \frac{1}{k\hspace{-.18cm}/ - m}\gamma ^\nu\frac{1}{ k'\hspace{-.26cm}/ - m} e^{- i\left(q+k-k'\right).x}e^{- i\left( -k - q +k' \right).y} \right]\varepsilon _\nu .
        \end{eqnarray}
        Integrating over $y$ and $k'$, the RHS gives,
        \begin{equation}
        {\varepsilon ^ * _\mu }\left[ { i{e^2}\int {\frac{{{d^d}k}}{{{{\left( {2\pi } \right)}^4}}}{\gamma ^\mu }\frac{1}{{k\hspace{-.18cm}/ - m}}{\gamma ^\nu }\frac{1}{{q\hspace{-.18cm}/ +k\hspace{-.18cm}/ - m}}} } \right]{\varepsilon _\nu }.
        \end{equation}
        By simple calculations we finally have,
        \begin{equation}
        \int d ^d y \widetilde{A}^{*}_{\mu }(i\Pi_2^ {\mu \nu})\widetilde{A}_\nu =\varepsilon ^{*}_{\mu }    \frac{{{-ie^2}}}{{6{\pi ^2}\epsilon }} (g^{\mu\nu}k^2-k^\mu k^\nu)   {\varepsilon _\nu }+O(\epsilon^0).
        \end{equation}
        Inserting the above calculation in Eq. (\ref{delta3}) and using $\widetilde{A}^{*}_{\mu}(x)\widetilde{A}_\nu(x)=\varepsilon^{*}_{\mu}\varepsilon_{\nu}$ we derive,
        \begin{eqnarray}
        {\delta _3}&=& \frac{{{-ie^2}}}{{6{\pi ^2}\epsilon }}\frac{{{\varepsilon^{ *}_{ \mu }}\left( {{g^{\mu \nu }}{k^2}-k^\mu k^\nu } \right){\varepsilon_\nu }}}{{{\widetilde{A}^{ *}_{ \mu }}[  -i\left( g^{\mu \nu } \left(- k^2\right)+k^\mu k^\nu \right)]{\widetilde{A}_\nu }}} +O(\epsilon^0)\nonumber\\&=&  - \frac{{{e^2}}}{{6{\pi ^2}\epsilon }}+O(\epsilon^0),
        \end{eqnarray}
        which is in accordance with Eq. (\ref{delta33}).
        
        For the  last counterterm, $\delta_1$, the numerator in Eq. (\ref{delta1}) can be rewritten as
        
        \begin{eqnarray}
        &&\int {d^d}z{d^d}y \, \overline \psi(z)   { {\delta\Gamma^\mu}} (x,y,z)\psi(y)= - {e^2}\int {{d^d}z {{d^d}y \ {\overline\psi}\left( {p',z} \right){\gamma ^\alpha }S\left( {z,x} \right){\gamma ^\mu}S\left( {x,y} \right){\gamma ^\beta }\psi \left( {p,y} \right){D_{\alpha\beta }}\left( {y,z} \right)} }
        \nonumber\\
        &&\hspace{2cm}= {e^2}\int d^dz {d^d}y  \overline u(p') \left[{e^{ip'.z}}\int \frac{{{d^d}k}}{{{{\left( {2\pi } \right)}^d}}} \frac{{{d^d}k'}}{{{{\left( {2\pi } \right)}^d}}} \frac{d^dk''}{{\left( {2\pi } \right)}^d}\right.\nonumber\\
        &&\hspace{4cm}\left.\times{\gamma ^\alpha }\frac{{{e^{ - ik'.\left( {z - x} \right)}}}}{{k'\hspace{-.25cm}/ - m}}{\gamma ^\mu}\frac{{{e^{ - ik.\left( {x - y} \right)}}}}{{k\hspace{-.18cm}/ - m}}{\gamma ^\beta }{e^{ - ip.y}}\frac{-ig_{\alpha\beta}}{k''^2}e^{- ik''.\left( {z - y} \right)}  \right]u(p)\nonumber\\
        &&\hspace{2cm}= -i{e^2}\overline u(p')\int \frac{{{d^d}k}}{{{{\left( {2\pi } \right)}^d}}} \frac{{{d^d}k'}}{{{{\left( {2\pi } \right)}^d}}} \frac{{{d^d}k''}}{{{{\left( {2\pi } \right)}^d}}}{\gamma ^\alpha }\frac{1}{{k'\hspace{-.25cm}/ - m}}{\gamma ^\mu}\frac{1}{{k\hspace{-.18cm}/ - m}}{\gamma ^\beta }\frac{{  {g_{\alpha\beta }}}}{{{{k''}^2}}}\nonumber\\
        &&\hspace{4cm}\times\left( {2\pi } \right)^{2d}{\delta^{(d)}}\left( {k + k'' - p} \right){\delta^{(d)}}\left( {p' - k' - k''} \right){e^{i\left( {k' - k} \right).x}}   u(p).
        \end{eqnarray}
        Taking integral of $k^{'}$ and $k^{''}$ yields,
        \begin{eqnarray}
        \nonumber \int {d^d}z{d^d}y \, \overline \psi(z)   { {\delta\Gamma^\mu}} (x,y,z)\psi(y)&=&-i{e^2}\overline u(p') \left[\int {\frac{{{d^d}k}}{{{{\left( {2\pi } \right)}^d}}}{\gamma ^\alpha }\frac{1}{{k'\hspace{-.25cm}/ - m}}{\gamma ^\mu}\frac{1}{{k\hspace{-.18cm}/ - m}}{\gamma ^\beta }\frac{{ - {g_{\alpha\beta }}}}{{{{\left( {p - k} \right)}^2}}}{e^{i\left( {p' - p} \right).x}}} \right]u(p)
        \\&=&\overline u(p')\left[ \frac{-e^2}{8\pi^2\epsilon}\gamma^\mu e^{i(p'-p).x}\right ]u(p) +O(\epsilon^0).
        \end{eqnarray}
        Replacing this result in Eq. (\ref{delta1}) we find,
        \begin{eqnarray}
        \delta_1\gamma ^{\mu}&=&\int {d^d}z {d^d}y\frac{\overline \psi(y)   { {\delta\Gamma^\mu}}(x,y,z) \psi(x) }{\overline \psi(x) \psi(x)}\Bigg|_{x=x_0}=\frac{\overline u(p')\left[ \frac{-e^2}{8\pi^2\epsilon}\gamma^\mu e^{i(p'-p).{x_0}}\right ]u(p)}{\overline u(p') u(p)e^{i(p'-p).{x_0}}}+O(\epsilon^0)\nonumber\\
        &\Rightarrow& \delta_1=-\frac{e^2}{8\pi^2\epsilon}+O(\epsilon^0),
        \end{eqnarray}
        which is again in complete agreement with Eq. (\ref{delta11}). This counterterm is equal to $\delta_2$ as it should be, due to the Ward identity. Consequently, up to order $\alpha$, we show that our counterterms in position space are equal to the usual terms derived in momentum space. Obviously, the results, in this case, do not depend on the special point $x_0$ where our renormalization conditions are imposed, manifesting the translational invariance of this problem.
        \section{Conclusions}\label{sec 5}
        Ultraviolet infinities of QED theory are basically due to three divergent Feynman diagrams: vertex correction, vacuum polarization and electron self-energy. Using renormalization program, in free space with translational symmetry, these infinities  are controlled by four counterterms which are generally derived in momentum space. However, if the translational invariance of the system is broken strongly then the momentum is no longer a good quantum number. Renormalization procedure in configuration space can be applied for such a situation, for example, in problems with a nontrivial BC or a nonzero background which cannot be treated as small perturbations.  In this paper, we have done the renormalization in real space in the presence of nontrivial BC and derived the form of four counterterms up to order $\alpha$. Systematic treatment of the renormalized perturbation theory after imposing renormalization conditions leads us to $x$-independent  counterterms which directly indicate the dependency on the BCs of the fermion and photon fields. Finally, as a particular case, our results have been compared with those obtained in free space and we have shown the equivalence in the two cases is guaranteed, up to order $\alpha$.

\end{document}